\documentclass[letterpaper,11pt]{article}
\pdfoutput=1

\usepackage{jheppub}
\usepackage{multirow}

\usepackage{subcaption}
\usepackage[countmax]{subfloat}
\usepackage{amssymb}
\usepackage{amsmath}
\usepackage{color}
\usepackage{graphicx}
\usepackage{verbatim}
\usepackage{amsthm}
\usepackage{slashed}
\usepackage{hyperref}

\def\cI{\mathcal{I}}

\def\cO{\mathcal{O}}

\def\nn{{\nonumber}}
\def\mcdot{\!\cdot\!}

\def\sss{\scriptscriptstyle}

\newcommand{\df}{\mathrm{d}}

\newcommand{\e}{\epsilon}

\newcommand{\zero}{{(0)}}
\newcommand{\one}{{(1)}}
\newcommand{\two}{{(2)}}

\newcommand{\as}{\alpha_{\scriptscriptstyle S}}
\def\mw{m_{\scriptscriptstyle W}}
\def\dd{\mathrm{d}}

\preprint{MIT-CTP/4911}

\title{Differential Distributions for t-channel Single Top-Quark Production and Decay at Next-to-Next-to-Leading Order in QCD}

\author{Edmond L. Berger$^2$, Jun Gao$^1$, Hua Xing Zhu$^{3,4}$}

\affiliation{$^1$INPAC, Shanghai Key Laboratory for Particle Physics and Cosmology, School of
Physics and Astronomy, Shanghai Jiao-Tong University, Shanghai 200240, China}
\affiliation{$^2$High Energy Physics Division, Argonne National Laboratory, Argonne, Illinois 60439, USA}
\affiliation{$^3$Center for Theoretical Physics, Massachusetts Institute of Technology, Cambridge, MA 02139, USA}
\affiliation{$^4$Zhejiang Institute of Modern Physics, Department of Physics, Zhejiang University, Hangzhou 310027, China}

\emailAdd{berger@anl.gov}
\emailAdd{jung49@sjtu.edu.cn}
\emailAdd{zhuhx@zju.edu.cn}

\abstract{
We present a detailed phenomenological study of the
next-to-next-to-leading order~(NNLO) QCD corrections for $t$-channel single top
(anti-)quark production and its semi-leptonic decay at the CERN Large Hadron Collider (LHC).
We find the NNLO corrections for the total inclusive rates at the LHC with different center of mass energies 
are generally smaller than the NLO corrections, indicative of improved convergence.  However,
they can be large for differential distributions, reaching a level of $10\%$ or 
more in certain regions of the transverse momentum distributions of the top (anti-)quark and the pseudo-rapidity 
distributions of the leading jet in the event.  In all cases the perturbative hard-scale uncertainties are
greatly reduced after the NNLO corrections are included.  We also show a comparison of
the normalized parton-level distributions to recent data from the 8 TeV measurement of the 
ATLAS Collaboration.  The NNLO corrections tend to shift the theoretical predictions
closer to the measured transverse momentum distribution of the top (anti)-quark.
Importantly, for the LHC at 13 TeV, we present NNLO cross sections in a fiducial
volume with decays of the top quark included.
}

\begin{document} 

\maketitle

\section{Introduction}
\label{sec:introduction}

The top quark ($t$) is the heaviest particle in the standard
model~(SM).   To date, it has been observed at hadron colliders
only through $t \bar{t}$ pair production or in single production.
Single top quark production provides a great
opportunity to directly probe the electroweak $Wtb$ vertex, which is
otherwise difficult to measure. There are three single-production channels: 
the $t$-channel through the exchange of a spacelike $W$ boson, the
$s$-channel through the exchange of a timelike $W$ boson, 
and associated production of $t$ with an on-shell $W$ boson.  Since
all three channels are directly connected to  
the $Wtb$ vertex, they can be used to measure the Cabibbo-Kobayashi-Maskawa (CKM) matrix element $V_{tb}$.
Besides, they can be used to extract the top-quark mass~\cite{Alekhin:2016jjz,CMS:2016fdm}
or to constrain the ratio of $u$-quark to $d$-quark parton
distributions~\cite{1404.7116,Alekhin:2015cza,Berger:2016oht}.
Single top-quark production is also sensitive to physics beyond the
SM~\cite{hep-ph/0007298}, e.g., modified structure of $Wtb$ vertex,
new gauge bosons or new
heavy quarks, and top-quark flavor-changing neutral current, and so forth.

%%%%%%%%%%%%%%%%%%%%%%%%%%%%%
% Experimental status
%%%%%%%%%%%%%%%%%%%%%%%%%%%%%

At a hadron collider such as the Fermilab Tevatron and CERN LHC, the
dominant mechanism for single top-quark production is through
$t$-channel exchange of a $W$ boson. This process was first observed at
the Tevetron~\cite{0903.0885,0903.0850}.  At the LHC, the $t$-channel cross
section has been measured by the ATLAS and CMS collaborations at $\sqrt{S}
= 7$ TeV~\cite{1106.3052,1205.3130,1209.4533,1406.7844}, $\sqrt{S} =
8$ TeV~\cite{1403.7366,1702.02859}, and $\sqrt{S} = 13$
TeV~\cite{1609.03920,1610.00678}. Recently, differential distributions
and fiducial cross section have also been measured~\cite{1702.02859}.  The CKM matrix element $V_{tb}$ and
the structure of the $Wtb$ vertex have been probed by ATLAS and
CMS~\cite{1403.7366,1510.03764,1702.08309}. The polarization of top quark in
$t$-channel production has also been measured~\cite{1511.02138}.

%%%%%%%%%%%%%%%%%%%%%%%%%%%
% Previous NLO results
%%%%%%%%%%%%%%%%%%%%%%%%%%%

Significant efforts have been made to improve the theoretical description of single top quark production. 
The next-to-leading order (NLO) QCD corrections in the 5-flavor scheme are calculated 
in Refs.~\cite{NUPHA.B435.23,hep-ph/9603265,hep-ph/9705398,hep-ph/9807340,hep-ph/0102126,
hep-ph/0207055,Sullivan:2004ie,hep-ph/0510224,1007.0893,1012.5132}.
The NLO calculation in the 4-flavor scheme is carried out in Ref.~\cite{0903.0005}.  
Full NLO corrections including top quark leptonic decay are studied within  
the on-shell top-quark approximation~\cite{hep-ph/0408158,hep-ph/0504230,1012.5132} and beyond~\cite{1102.5267,1305.7088,1603.01178}.  Code for fast numerical evaluation at NLO is 
provided in Ref.~\cite{1406.4403}.
Soft gluon resummation is considered in
Refs.~\cite{1010.4509,1103.2792,1210.7698,1510.06361}. Matching NLO
calculations to parton showers is done in the framework of POWHEG and MC@NLO
Refs.~\cite{hep-ph/0512250,0907.4076,1207.5391,1603.01178}. For experimental analyses
at the LHC, predictions from POWHEG or MC@NLO are used for
modeling of the signal process in unfolding to parton level cross
sections, as well as for comparison of data and theory.  
The cross sections from either measurement or prediction can have a theoretical
uncertainty of about $5-10\%$~\cite{1702.02859}.  Predictions incorporating
further higher-order or logarithmic corrections are desirable for precision
measurements.   

%%%%%%%%%%%%%%%%%%%%%%%%%%%
% Introduce the NNLO results
%%%%%%%%%%%%%%%%%%%%%%%%%%%

Next-to-next-to-leading order (NNLO) QCD
corrections with a {\em stable} top quark are calculated in
Refs.~\cite{1404.7116,Berger:2016oht}, with neglect of certain subleading contributions in color,
namely in the structure-function approximation. The calculation of
Ref.~\cite{Berger:2016oht} also includes 
the top-quark leptonic decay at NNLO within the
on-shell top-quark approximation. Thus for the first time a realistic parton-level
simulation at NNLO is available. The NNLO QCD effects
on experimental fiducial cross sections at 13 TeV
are reported in Ref.~\cite{Berger:2016oht}. The corrections
are found to be large, both from production and
decay, owing mostly to the jet-veto condition in the definition of the fiducial volume. 
In this paper we provide further elaboration of the methods and numerical results of 
our NNLO calculation.   We present NNLO results for the LHC at 7, 8, and 14 TeV,  in 
particular the total inclusive cross sections and differential distributions
with a stable top quark.     

The rest of our paper is organized as follows. In
Sec.~\ref{sec:detail-calculation}, we expand upon our NNLO 
calculation presented in Ref.~\cite{Berger:2016oht} and its validation.  
In Sec.~\ref{sec:incl-cross-sect}, we present our predictions for the total
inclusive cross sections and the differential cross sections with a stable top-quark in
the final state.  Sec.~\ref{sec:fiduc-cross-sect} provides results on fiducial cross sections and
distributions for which the top-quark decay is included through NNLO, enabling a more refined 
comparison with data.   Readers interested 
principally in comparisons with experiment may chose to bypass Sec.~\ref{sec:detail-calculation} 
on a first reading of this paper.  Finally our summary and conclusions are presented in 
Sec.~\ref{sec:conclusions}.
Our results show that the NNLO QCD corrections are large in certain regions of the 
differential distributions as well as for fiducial cross sections with jet veto selections.  They 
stabilize the theoretical predictions, with residual scale variations of about one percent.
The NNLO predictions provide an improved description of the transverse momentum 
distribution of the top quark measured by ATLAS collaboration.
%  

%%%%%%%%%%%%%%%%%%%%%%%%%%%%%

\section{Theoretical Framework}
\label{sec:detail-calculation}

We describe in this section our calculation of single top-quark
production and decay at hadron colliders through NNLO. The calculation for
single antitop-quark production follows the same line of reasoning. Some of the results have been
presented in our previous publication~\cite{Berger:2016oht}.

\begin{figure}[ht]
  \centering
  \includegraphics[width=0.4\textwidth]{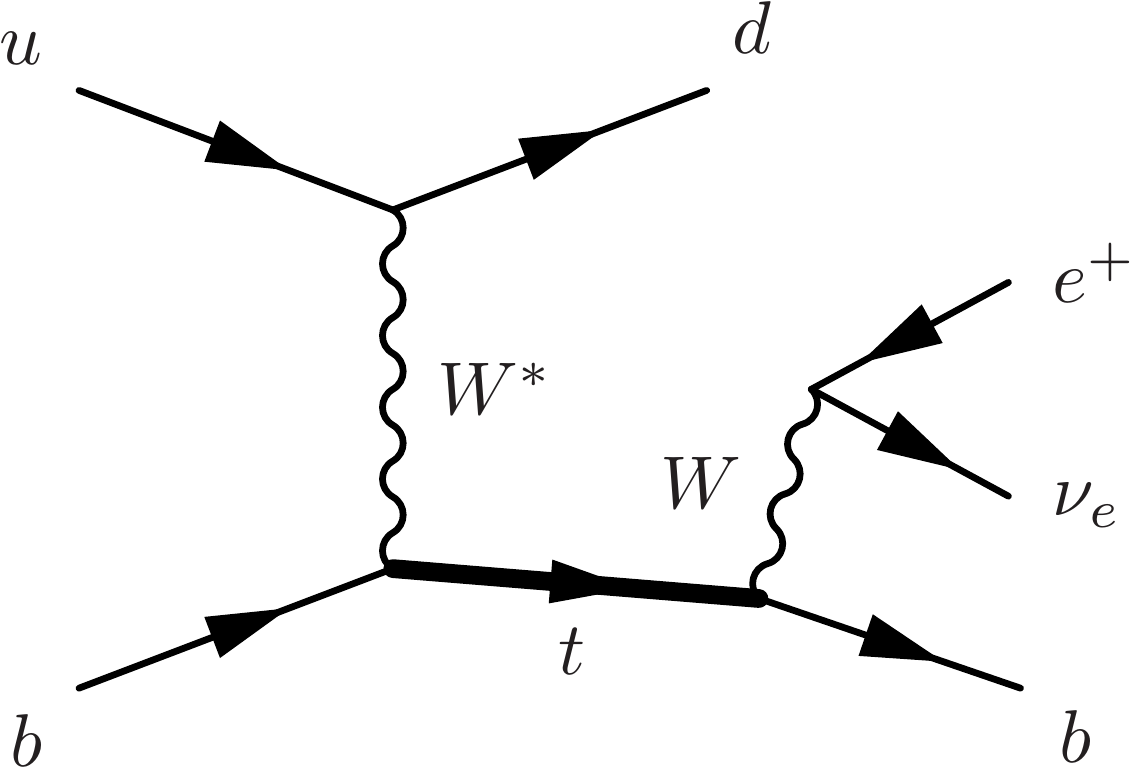}
  \caption{The LO Feynman diagram for single top-quark production and
    decay at hadron colliders. Top quark is represented by a thick
    line. We show only one partonic channel in this figure.}
  \label{fig:born}
\end{figure}

The LO Feynman diagram for the process under consideration is depicted
in Fig.~\ref{fig:born}.  We first discuss the approximations we 
employed to make the calculation feasible, namely, the on-shell
top quark approximation~\cite{hep-ph/9309234,hep-ph/9302311} and the structure-function 
approximation~\cite{hep-ph/9705398}.  Thanks
to these approximations, the calculation effectively factors into
three separate calculations with much simpler structure.  We then
present detailed formulas for these three simpler calculations.  
We discuss the validation of our calculation toward the end of
this section. 

\subsection{On-shell top-quark approximation and structure-function
  approximation}
\label{sec:on-shell}

In our calculation, we neglect interference between real radiation
from the single top production stage and the top-quark decay stage. We also
neglect a term in which there is a virtual gluon connecting the production
and decay stages.  This approximation is known as the on-shell top-quark
approximation, {\it i.e.}, the top quark is on its mass shell in all
the diagrams when considered as an external state.  For a generic
inclusive enough infrared-safe observable, the omitted corrections are suppressed by
the width of the top quark, $\Gamma_t/m_t$~\cite{hep-ph/9309234,hep-ph/9302311}.  In the SM, top quark has a
relatively small width but a large mass, $\Gamma_t/m_t <
1\%$, an the approximation should provide an excellent representation of the full prediction.   The on-shell approximation has been 
used in $t$-channel
single top production by different
groups~\cite{NUPHA.B435.23,hep-ph/9705398,hep-ph/0207055,hep-ph/0504230,hep-ph/0408158,1012.5132}
at NLO, and recently at NNLO~\cite{1404.7116,Berger:2016oht}. Effects
beyond the on-shell approximation have been explored only at NLO thus
far~\cite{hep-ph/9603265,1007.0893}, owing to the complexity of the
calculation.

Through NNLO, the on-shell top-quark approximation can be written as
\begin{align}
  \label{eq:2}
    \sigma^{\rm LO} = &\,  \frac{1}{\Gamma^\zero_t} \dd \sigma^\zero
                        \otimes \dd
                      \Gamma^\zero_t
\nn\\
\delta\sigma^{\rm NLO} = & \,  \frac{1}{\Gamma^\zero_t}\left[\dd
                           \sigma^\one \otimes \dd
                     \Gamma^\zero_t+  \dd \sigma^\zero \otimes\left(\dd
                     \Gamma^\one_t -
                     \frac{\Gamma^\one_t}{\Gamma^\zero_t} \dd \Gamma^\zero_t\right)\right]
\nn\\
\delta\sigma^{\rm NNLO} = & \,  \frac{1}{\Gamma^\zero_t}\left[ \dd
                            \sigma^\two \otimes \dd
                     \Gamma^\zero_t
+ \dd \sigma^\one \otimes \left(\dd
                     \Gamma^\one_t-\frac{\Gamma^\one_t}{\Gamma^\zero_t}\dd
                            \Gamma^\zero_t\right)
\right.
\nn\\
&
\left.
 + \dd \sigma^\zero \otimes \left(\dd
                     \Gamma^\two_t
 - \frac{\Gamma^\two_t}{\Gamma^\zero_t} \dd \Gamma^\zero_t  - 
\frac{\Gamma^\one_t}{\Gamma^\zero_t} \left(\dd \Gamma^\one_t-\frac{\Gamma^\one_t}{\Gamma^\zero_t}\dd \Gamma^\zero_t\right)\right)\right],
\end{align}
where $\Gamma_t^{(0),(1),(2)}$ and $\sigma^{(0),(1),(2)}$ denote the
Born,
$\cO(\as)$, and $\cO(\as^2)$ top-decay width and production cross
section, respectively. In Eq.~\eqref{eq:2} we have expanded the QCD
corrections to both production and decay to the same order
consistently. Equation~\eqref{eq:2} can be used for a fully differential
calculation.  After integrating over phase space, one reproduces the
inclusive production cross section at a given order, as expected.  
For a correct treatment of spin correlations, the
production cross section $\dd\sigma$ and the decay width $\dd\Gamma_t$
must be calculated for an on-shell polarized top quark. The symbol
$\otimes$ denotes the appropriate summation over polarization.

\begin{figure}[ht]
  \centering
  \begin{subfigure}[b]{0.3\textwidth}
    \centering
    \includegraphics[height=0.2\textheight]{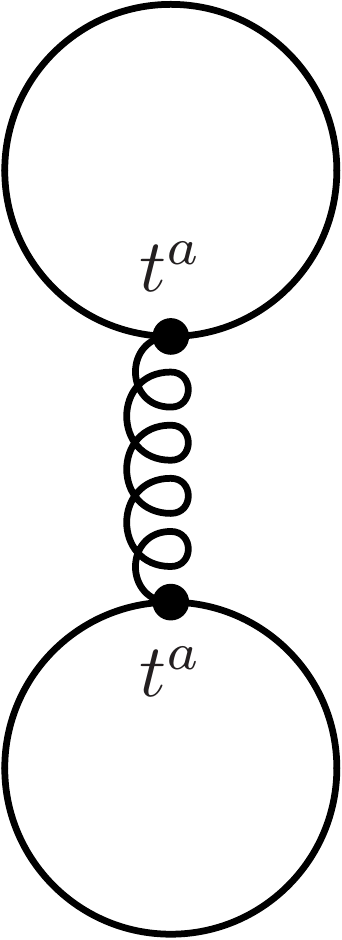}
\caption{\label{fig:2a}}
  \end{subfigure}
  \begin{subfigure}[b]{0.3\textwidth}
    \centering
    \includegraphics[height=0.2\textheight]{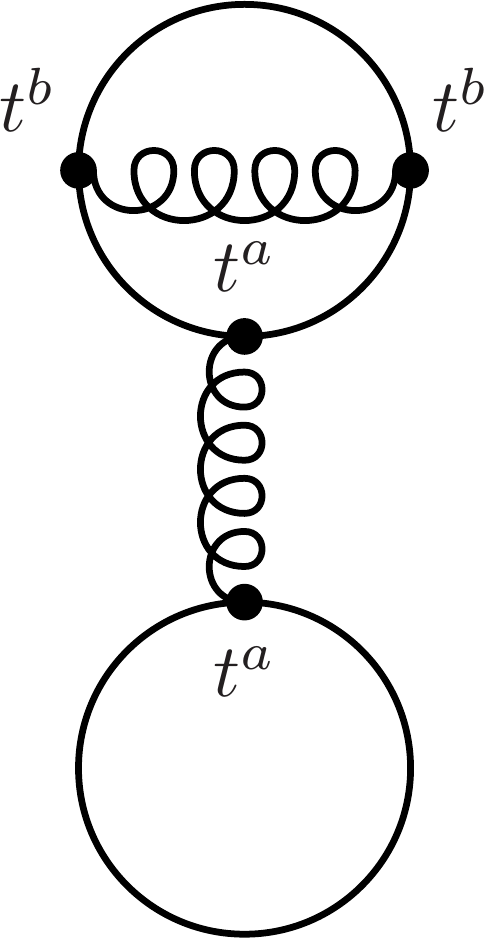}
\caption{\label{fig:2b}}
  \end{subfigure}
  \begin{subfigure}[b]{0.3\textwidth}
    \centering
    \includegraphics[height=0.2\textheight]{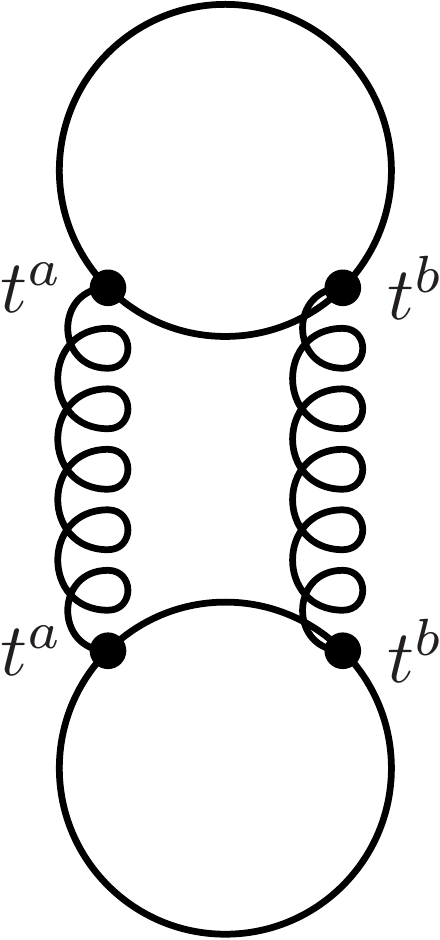}
\caption{\label{fig:2c}}
  \end{subfigure}
\caption{Examples of the color component of NLO and NNLO Feynman diagrams for
  $t$-channel single top-quark production.  Both virtual and real
  diagrams can be represented in this form. The lower loop represents
  the heavy-quark line, whereas the upper loop represents the
  light-quark line. \label{fig:2}}
\end{figure}

Even with the on-shell top-quark approximation, the full NNLO QCD
corrections to the production stage remain very difficult. For
example, the full two-loop diagrams involve four different scales, the
Mandelstam variables $s$ and $t$, the top quark mass $m_t$, and the $W$ boson
mass $\mw$.  A full two-loop amplitude of this complexity has not
been obtained yet, either analytically or numerically, though
interesting progress has been made~\cite{1409.3654,1611.01087}. To
bypass this complexity, we adopt the structure-function approximation~\cite{hep-ph/9705398},
namely, we systematically neglect virtual and real radiation interference between the
light quark line and the heavy quark line.  These effects vanish exactly at
NLO for squared amplitudes, owing to the traceless-ness of Gell-Mann
matrices. This result can be seen from the color component of the
real or virtual Feynman diagram for the
NLO squared amplitudes with interference between the light and
the heavy quark line, Fig.~\ref{fig:2a}, which is proportional to
$\mathrm{Tr}[t^a] \mathrm{Tr}[t^a] =0$. This result is true even for part of the
NNLO diagrams, as long as there is only one gluon exchanged between the 
light and the heavy quark line, such as the diagram in
Fig.~\ref{fig:2b}.  However, it ceases to be true for the
diagrams with two gluons exchanged between the light and the heavy quark
line, such as the diagram in Fig.~\ref{fig:2c}.  Such a diagram has a
color factor $\mathrm{Tr}[t^a t^b] \mathrm{Tr}[t^a t^b] = (N_c^2 -
1)/4$, which is suppressed by a factor of $1/N_c^2$ compared with
those without light quark and heavy quark line interference.  In the
structure-function approximation, such diagrams are neglected, in both 
virtual and a real corrections.  These contributions are gauge
invariant and IR finite by themselves, justifying the
structure-function approximation.  This approximation has been
employed in the previous NNLO calculation for single top-quark
production~\cite{1404.7116,Berger:2016oht}.  We note finally that the separation 
of single top-quark production into $t$-channel and $s$-channel terms could
be ambiguous at NNLO, because there exist NNLO contributions which are the 
interference between one-loop $s$-channel and $t$-channel diagrams. These
contributions fall into the class of two-gluon exchange diagrams in
Fig.~\ref{fig:2c}.  These contributions are not present in the structure-function
approximation, consistent with the use of $t$-channel in the
title of this work.

\begin{figure}[ht]
  \centering
  \includegraphics[width=0.4\textwidth]{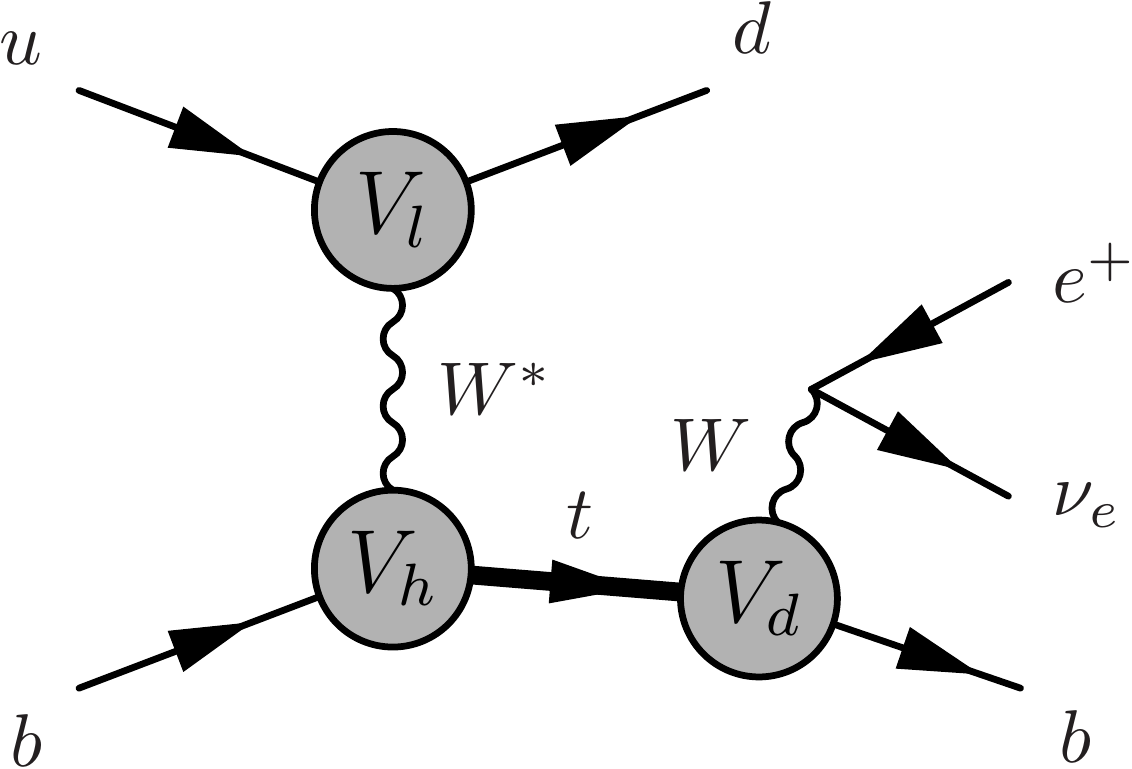}
  \caption{Schematic diagram for $t$-channel single top-quark
    production at hadron colliders in the on-shell top quark
    approximation and the structure-function approximation. The full QCD
    corrections are factored into three different parts with these approximations.}
  \label{fig:3}
\end{figure}

The on-shell top quark approximation and structure-function
approximation can be summarized schematically in
Fig.~\ref{fig:3}.   Owing to these approximations, the full QCD
corrections are factored into a piece describing the decay of the top quark,
$V_d$, DIS-like production of the top quark, $V_h$, and the DIS-like
production of a light jet, $V_l$.  In the remainder of this section, we
shall discuss the QCD corrections to each of these three parts
separately. 

\subsection{QCD corrections for decay of the top quark}
\label{sec:qcd-corrections-top}

In this subsection we discuss the calculation of the fully differential
semi-leptonic decay rate of a top quark through NNLO.  These results were first 
presented in Ref.~\cite{1210.2808}.  We provide more details here. 

A typical QCD calculation beyond LO consists of virtual corrections,
real emission corrections, and also mixed real-virtual corrections in the case
of NNLO or beyond. A well-known feature of an on-shell perturbative QCD
calculation is that individual piece of the higher order QCD
corrections contains infrared divergences.  The divergences cancel only in the sum
of different contributions for infrared-safe observables.   For example,
the virtual corrections contain explicit infrared poles as a result of soft or
collinear modes of the loop integrals.  For real emission corrections, infrared
poles result from integrating real radiation terms over unresolved phase
space regions.  These implicit poles in the corrections prevent a
naive Monte-Carlo integral over the phase space in four space-time
dimensions.   A successful higher order QCD calculation requires the use of
procedures to regulate the infrared singularities in the phase
space integral. 

In Ref.~\cite{1210.2808}, the phase space slicing method was employed to
regulate the infrared singularities in the phase space integral. The
idea of phase space slicing method is simple. For any infrared-safe
observable $O$, the differential distribution can be written as
\begin{align}
  \label{eq:3}
  \frac{\dd\sigma}{\dd O} = & \int^{\rho_{\rm cut}}_0 \dd \rho \, \frac{\dd^2 \sigma}{\dd O
  \, \dd \rho} + \int^{\rho_{\rm max}}_{\rho_{\rm cut}} \dd \rho \, \frac{\dd^2 \sigma}{\dd O
  \, \dd \rho} 
\nn\\
= & \left. \frac{\dd\sigma}{\dd O} \right|_{\rm unres.} +
    \left. \frac{\dd\sigma}{\dd O} \right|_{\rm res.} \,,
\end{align}
where we have introduced a resolution variable $\rho$, and split the
integral into an unresolved part, the first term on the RHS, and an
resolved part, the second term on the RHS.  There is no canonical
definition for the resolution variable. The only requirement is that
$\rho \to 0$ in the unresolved limit.  In the resolved part, it is
demanded that no phase space singularity be presented in the matrix
element, and the integral can be performed in four dimensions using
a Monte-Carlo method.  The key idea of the phase space
slicing method is that the QCD matrix element in the unresolved part can
be approximated by the soft or
collinear singular limit of the corresponding matrix element, known to
have a universal factorized form.   Owing to the simplicity of the QCD
matrix element and phase space in the soft or collinear limit, it is sometimes possible to
perform the  unresolved phase space integral analytically.  The
infrared poles from the unresolved phase space integral can then be extracted
in analytic form and cancelled against the corresponding infrared poles from
the virtual corrections.  The approximation in the unresolved part
introduces an $\cO(\rho_{\rm cut} \ln^k \rho_{\rm cut})$ error compared with
the true calculation.  To reduce this error as much as possible, and not modify 
the physical observable $O$ significantly, it is
desirable to choose a small cut-off $\rho_{\rm cut}$ for the
resolution variable.  However, a small $\rho_{\rm cut}$ will
also lead to a very steep integrand for the resolved part, and
therefore potentially large Monte-Carlo integration 
uncertainty.  Within the N-jettiness subtraction
formalism~\cite{1504.02131,1505.04794}, progress has been made 
recently in reducing the 
analytic error in the unresolved part by incorporating the leading
logarithms of the power suppressed terms~\cite{1612.00450,1612.02911}. 

In Ref.~\cite{1210.2808}, the resolution variable is chosen as the
inclusive jet mass, normalized to the top-quark mass,
\begin{align}
  \label{eq:4}
  \tau_d = \frac{(\sum_i p_i)^2}{m_t^2} \,,
\end{align}
where the sum runs over all final-state QCD partons $p_i$.  We use a subscript
$d$ to denote the resolution variable chosen in the decay
calculation.  It is easy to see that $\tau_d$ satisfies the requirement
for a good resolution variable for phase space slicing, namely $\tau_d
\to 0$ when all the final-state QCD partons are either soft or
collinear with each other. 

We write the differential decay rate as
\begin{align}
  \label{eq:5}
  \frac{\dd \Gamma_t}{\dd O} = & \int^{\tau_{d,\rm cut}}_0 \dd \tau_d \, \frac{\dd^2 \Gamma_t}{\dd O
  \, \dd \tau_d} + \int^{\tau_{d,\rm max}}_{\tau_{d,\rm cut}} \dd \tau_d \, \frac{\dd^2 \Gamma_t}{\dd O
  \, \dd \tau_d} 
\nn\\
= & \left. \frac{\dd\Gamma_t}{\dd O} \right|_{\rm unres.} +
    \left. \frac{\dd\Gamma_t}{\dd O} \right|_{\rm res.} \,.  
\end{align}
The task is to compute the unresolved part using an approximated QCD
matrix element and phase space, and the resolved part using numerical
Monte-Carlo integral.  We stress that Eq.~\eqref{eq:5} holds for both
the polarized and the unpolarized decay rate.  Since our goal is to
combine production and decay at NNLO, we compute the polarized decay rate
in this work.  For decay of the $W$ boson we adopt the narrow-width
approximation. Therefore, we consider the decay of top quark to
an on-shell $W$ boson and a $b$ quark at LO, while keeping the full
polarization information for both the top quark and the $W$ boson.  
Because of the simple form of the resolution
variable, the integrand of the unresolved part can be written in a convenient
factorized form, up to error terms proportional to $\cO(\ln^k \tau_d)$
with $k_s \geq 0$,
\begin{align}
  \label{eq:6}
 \left.\frac{\dd^2 \Gamma_t}{\dd O \, \dd \tau_d} \right|_{\rm unres.}= \frac{\dd \Gamma_t^\zero}{\dd
  O} H_d(x,\mu) \int \dd m^2 \, \dd k_s \, J(m^2,\mu) S_d(k_s,\mu)
  \delta\left(\tau_d  - \frac{m^2 + 2 E_J k_s }{m_t^2} \right) +
  \cO(\ln^k \tau_d) \,,
\end{align}
where $x = \mw^2/m_t^2$ 
characterizes the LO decay kinematics, and $E_J = (m_t^2 - \mw^2)/(2
m_t)$ is the energy of the $b$ jet at LO.
Such a factorization formula was originally discussed in inclusive
B decay in the end point
region~\cite{Korchemsky:1994jb,Akhoury:1995fp,Bauer:2003pi,Bosch:2004th}
using Heavy Quark Effective Theory~(HQET) and Soft-Collinear Effective
Theory~(SCET)~\cite{hep-ph/0005275,hep-ph/0011336,hep-ph/0109045,hep-ph/0202088}.  
The same factorization
formula can be used in top-quark decay since the
observable is very similar, as long as its use is restricted to the perturbative
region. Equation~\eqref{eq:6} indicates that in the unresolved region, the
kinematic distributions for $O$ follow exactly those at LO. The
normalization is determined in a factorization-friendly form in terms
of a hard function $H_d(x,\mu)$, a jet function $J(m^2,\mu)$, and a
heavy-quark decay soft function $S_d(k_s,\mu)$. The universality of
infrared dynamics of QCD implies that these functions for top decay
can also be extracted from those for inclusive B decay.  Note that in
Eq.~\eqref{eq:6}, the polarization information of the top quark and the $W$
boson is encoded in the LO decay rate and the hard function only. 

\begin{figure}[ht]
  \centering
  \begin{subfigure}[b]{0.3\textwidth}
    \centering
    \includegraphics[height=0.15\textheight]{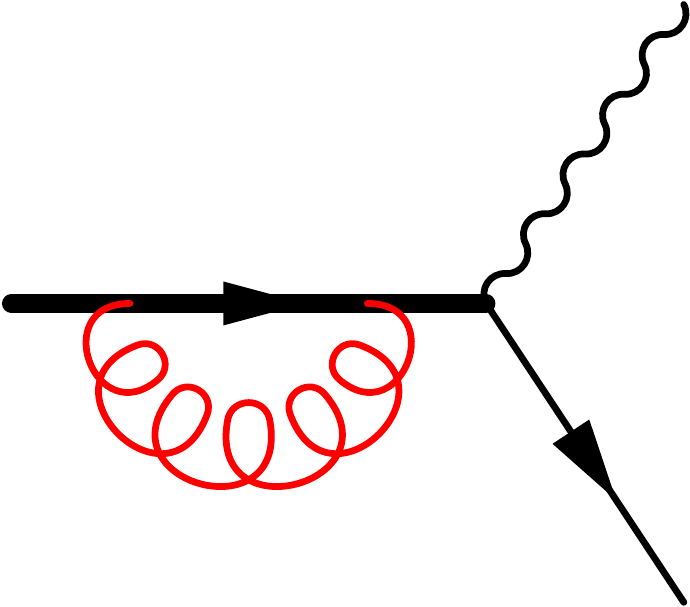}
\caption{\label{fig:4a}}
  \end{subfigure}
  \begin{subfigure}[b]{0.3\textwidth}
    \centering
    \includegraphics[height=0.15\textheight]{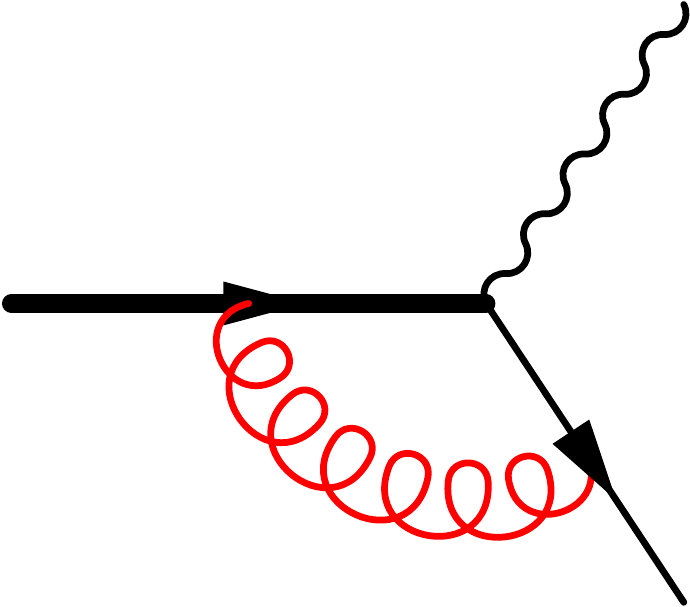}
\caption{\label{fig:4b}}
  \end{subfigure}
\caption{One-loop QCD form factor for heavy-to-light decay.\label{fig:4}}
\end{figure}

The hard function is related to the operator
resulting from matching the heavy-to-light QCD form factor onto HQET and
SCET. To leading power in the heavy quark limit and to all orders in
$\as$, the operator can be expanded in terms of three basis functions,
\begin{align}
  \label{eq:7}
  \cO_{tb} = C_1(x, \mu) \bar{\chi}_n \slashed{\varepsilon} (1 -
  \gamma_5) h + C_2(x, \mu) v \mcdot \varepsilon  \bar{\chi}_n (1 +
  \gamma_5) h + C_3(x,\mu) \frac{n \mcdot \varepsilon}{n \mcdot v}
  \bar{\chi}_n ( 1  + \gamma_5) h \,,
\end{align}
where $n^\mu = p_b^\mu/p_b^0$ and $v^\mu = p_t^\mu/m_t$ are the
four-velocity of the $b$ jet and the top quark at LO in QCD; $\chi_n$ is the
gauge-invariant collinear $b$ quark field; $h$ is the heavy top quark
field; and $\varepsilon$ is the polarization vector for the $W$ boson. The
Wilson coefficients $C_i(x,\mu)$ can be extracted from the QCD form
factor calculation. For example, at one-loop, the relevant diagrams are shown in
Fig.~\ref{fig:4}. 

The Wilson coefficients $C_i(x,\mu)$ have been calculated through two loops
for inclusive B decay~\cite{Bonciani:2008wf,Asatrian:2008uk,Beneke:2008ei,Bell:2008ws}.  The 
corresponding Wilson coefficients for
top-quark decay can be simply read off from these studies.  We
quote the results through $\cO(\as)$ below,
\begin{align}
  \label{eq:8}
  C_1(x,\mu) = & 1 + \frac{\as}{4\pi} C_F \left(-2
                 \ln^2\frac{\mu}{m_t}+4 \ln\frac{\mu}{m_t} \log
                 (1-x)-5 \ln\frac{\mu}{m_t}+2 \text{Li}_2(1-x)-2 \log ^2(1-x)
\right.
\nn\\
&
\left. -\frac{\log (1-x)}{x}+3 \log (1-x)+2 \log (1-x) \log (x)-\frac{5 \pi
   ^2}{12}-6\right) + \cO(\as^2)\,,
\\
C_2(x,\mu) = & 0 + \cO(\as^2) \,,
\\
C_3(x,\mu) = & \frac{\as}{4\pi} C_F \left(-\frac{2 \log
               (1-x)}{x^2}-\frac{2}{x}+\frac{4 \log (1-x)}{x}\right) +
               \cO(\as^2) \,.
\end{align}
We refer readers to
Refs.~\cite{Bonciani:2008wf,Asatrian:2008uk,Beneke:2008ei,Bell:2008ws}
for the full two-loop results\footnote{In our calculation, we use the
  result of Ref.~\cite{Asatrian:2008uk}, kindly provided to us by Ben
  Pecjak in a convenient computer readable form.}.

The hard function is defined as the squared matrix element of the
effective operator normalized to the Born level result,
\begin{align}
  \label{eq:9}
  H_d(x,\mu)  = \frac{ \left| \langle Wb | \cO_{tb} | t \rangle
  \right|^2}{ \underset{\as \to 0}{\lim}\left| \langle Wb | \cO_{tb} | t \rangle
  \right|^2} \,.
\end{align}

The soft function is defined as a vacuum matrix element of Wilson
loops, which is independent of the top-quark spin.  In a practical
calculation, they can be obtained by taking the eikonal limit of the
real corrections, with the insertion of a 
measurement function $\delta( k_s - k \mcdot n)$, where $k_s$ is the
total momentum of the soft radiation in the final state.  For instance, the
one-loop soft function is given by the integrals
\begin{align}
  \label{eq:10}
  S^\one_d(k_s,\mu) = \mu^{2 \e} \int \frac{\dd^{4 - 2 \e} k}{(2\pi)^{4 -  2
  \e}} (2\pi) \Theta(k^0) \delta(k^2) \delta( k_s - k \mcdot n) \left| \parbox[h]{0.18\textwidth}{
\includegraphics[width=0.15\textwidth]{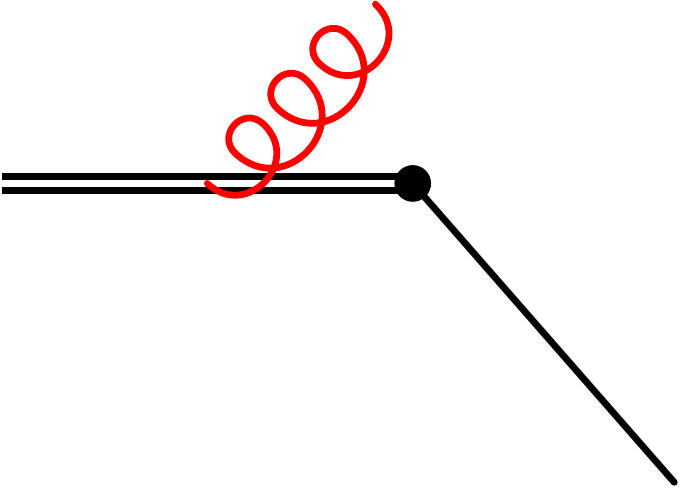} 
}
+ 
\parbox[h]{0.18\textwidth}{
\includegraphics[width=0.15\textwidth]{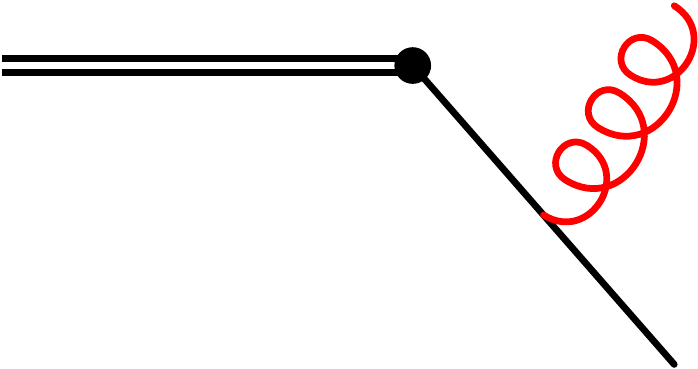} }
\right|^2 \,.
\end{align}
We use a double line to denote a timelike Wilson line, and a solid
real line to denote a lightlike Wilson line.  Note that the definition
for the soft function is not Lorentz invariant.  The violation of
Lorentz invariance comes only from the measurement function
$\delta(k_s - k \mcdot n)$.  The full two-loop soft
function for inclusive B decay in the rest frame of a B meson has been computed in
Ref.~\cite{Becher:2005pd}. The top quark decay soft function in the top quark rest
frame is exactly the
same as the B decay soft function, owing to universality of QCD amplitudes in the
soft limit. We quote the result for the soft function through one loop below,
\begin{align}
  \label{eq:11}
  S_d(k,\mu) = \delta(k) + \frac{\as}{4\pi} C_F \left( - 8 \left[
  \frac{\ln(k/\mu)}{k}\right]_\star^{[k,\mu]} - 4 \left[
  \frac{1}{k}\right]_\star^{[k,\mu]} - \frac{\pi^2}{6} \delta(k) \right) + \cO(\as^2) \,,
\end{align}
where the star distribution is defined as
\begin{align}
  \label{eq:12}
  \int^\mu_0 \dd k \, [f(k)]_\star^{[k,\mu]} g(k)  =   \int^\mu_0 \dd
  k \, f(k)( g(k) - g(0) ) \,.
\end{align}
We refer to Ref.~\cite{Becher:2005pd} for the full two-loop soft
function. 

The jet function is defined as the vacuum matrix element of the gauge
invariant collinear field with the insertion of a measurement function
$\delta(m^2 - p^2)$, where $p^2$ is the virtuality of the collinear
jet. In practice, it can be calculated by integrating the unintegrated
splitting function with the above mentioned delta function
inserted. At one-loop, the quark jet function is given by the
integral
\begin{align}
  \label{eq:13}
  J(m^2,\mu) = \mu^{2 \e} \int \frac{\dd^{4 - 2 \e} l}{(2\pi)^{4 -  2
  \e}} (2\pi) \Theta(l^0) \delta(l^2) \delta(m^2 - p^2 )\left| \parbox[h]{0.18\textwidth}{
\includegraphics[width=0.15\textwidth]{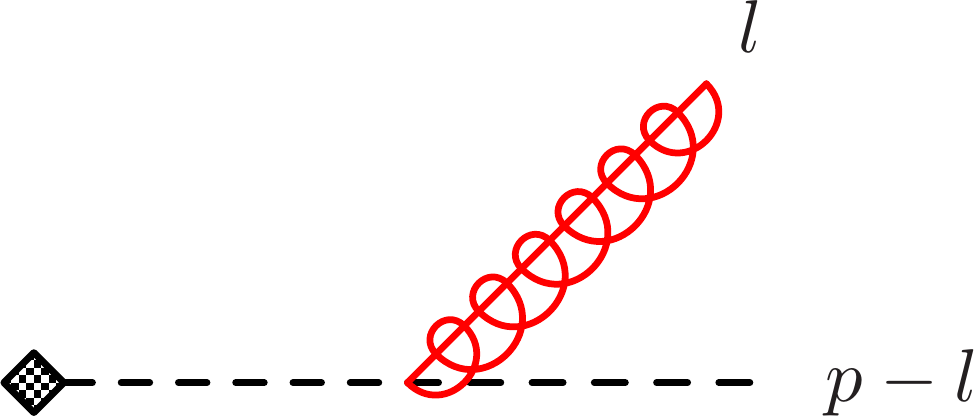} 
}
+ 
\parbox[h]{0.18\textwidth}{
\includegraphics[width=0.15\textwidth]{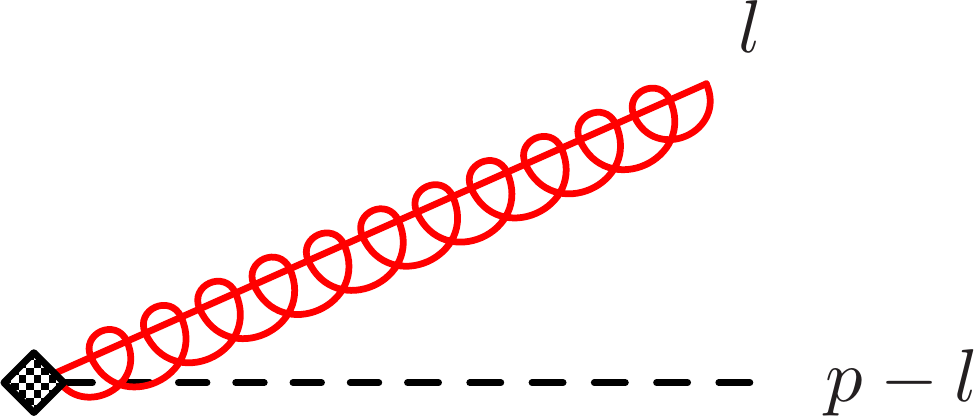} }
\right|^2 \,.
\end{align}
 We use a hatched diamond to denote collinear Wilson line in
 SCET. The jet function is completely factorized from the top quark,
 therefore also independent of top quark polarization. The
 overlap region of soft and collinear gluons is removed by the zero-bin
 subtraction procedure~\cite{hep-ph/0605001}. For the inclusive jet
 function the zero-bin subtraction term vanishes to all orders in
 $\as$. The one-loop jet function is 
 \begin{align}
   \label{eq:14}
   J(m^2,\mu) = \delta(m^2) + \frac{\as}{4\pi} C_F \left( 4
   \left[\frac{\ln(m^2/\mu^2)}{m^2}\right]_\star^{[m^2,\mu^2]} - 3
   \left[\frac{1}{m^2}\right]_\star^{[m^2,\mu^2]}  + (7 - \pi^2 ) \delta(m^2)\right)
 \end{align}
For this work, we need the full two-loop quark jet function, which was computed in Ref.~\cite{Becher:2006qw}.

After putting the one-loop hard, soft, and jet functions together, one can 
derive the one-loop prediction for the unresolved integrand,  
\begin{align}
  \label{eq:15}
   \left.\frac{\dd^2 \Gamma_t}{\dd O \, \dd \tau_d} \right|_{\rm
  unres.}= &  \frac{\dd \Gamma_t^\zero}{\dd 
  O} \left\{ \delta(\tau_d) + \frac{\as}{4 \pi} C_F \left[ - 4
             \left[\frac{\ln \tau_d}{\tau_d} \right]_+ +\left( 8 \ln(1-x) - 7 \right) \left[ \frac{1}{\tau_d} \right]_+ 
\right.\right.
\nn\\
&
\left.
\left.
+ \left( 4 \ln^2\frac{\mu}{m_t} - 8 \ln(1-x)
  \ln\frac{\mu}{m_t} + 10 \ln\frac{\mu}{m_t} - 4 \ln^2(1-x) + 4 \ln(1-x)
\right.
\right.
\right.
\nn\\
&
\left.
\left.
\left.
+ 7 - \frac{7 \pi^2}{6} + H_d^\one(x,\mu) \right) \delta(\tau_d)
\right]  \right\} + \cO(\ln^k \tau_d) \,,
\end{align}
where $[f(x)]_+$ is the usual plus distribution.  The $\mu$ dependence 
cancels completely at this order, once the one-loop hard function
is inserted. This is equivalent to the statement that the infrared
divergences have been cancelled between virtual and real corrections at
this order.  Note that up to power
corrections, the $\tau_d$ dependence of the unresolved integrand is
very simple and can be integrated out readily in Eq.~\eqref{eq:5}. 

The factorized form of Eq.~\eqref{eq:6} is very convenient for calculations 
at higher order. Indeed, the only ingredients needed for a NNLO
calculation of the unresolved part are the corresponding two-loop hard,
jet, and soft functions, which are available from the previous
precision study of inclusive B decay.  This is one of the advantages of
the phase space slicing method within the effective field theory
framework, namely the convenient organization of different
perturbative ingredients and the ease of recycling existing universal
functions.

For a small cut-off $\tau_{d,\rm cut}$, integration of the unresolved
distribution obtained from the factorization formula results in large
logarithmic dependence on the cut-off.  At NLO, the leading term scales as $\ln^2
\tau_{d,\rm cut}$ as is evident from Eq.~\eqref{eq:14}, whereas at NNLO 
it scales as $\ln^4 \tau_{d,\rm cut}$.  For sufficiently small cut-off, the large cut-off 
dependence is to be cancelled by the resolved contribution, up to Monte-Carlo
integration uncertainty.  The resolved contribution, as its name
suggests, is free of infrared singularities at NLO.  At NNLO, the
resolved contribution contains sub-divergences. These sub-divergences
cannot be resolved by our resolution variable $\tau_d$. They must be
cancelled using other methods.  Fortunately, the infrared structure of
sub-divergences is lower by one order in $\as$ than the unresolved
part.  For a NNLO calculation, we can use any existing subtraction
method to cancel the sub-divergences. In our calculation, we employ
the dipole subtraction formalism~\cite{hep-ph/9605323} with
appropriate massive dipole terms~\cite{Melnikov:2011qx} to remove the
sub-divergences.  We also need the one-loop amplitudes for top quark decay to
a $W$ plus two partons, and tree-level amplitudes for top decay to a $W$
plus three partons.  We extract the former from
Ref.~\cite{Campbell:2005bb}; for the latter we use
\texttt{HELAS}~\cite{Murayama:1992gi}.

\subsection{QCD corrections for production of a single top quark: heavy quark line}
\label{sec:qcd-corr-single}

\begin{figure}[ht]
  \centering
  \includegraphics[width=0.3\textwidth]{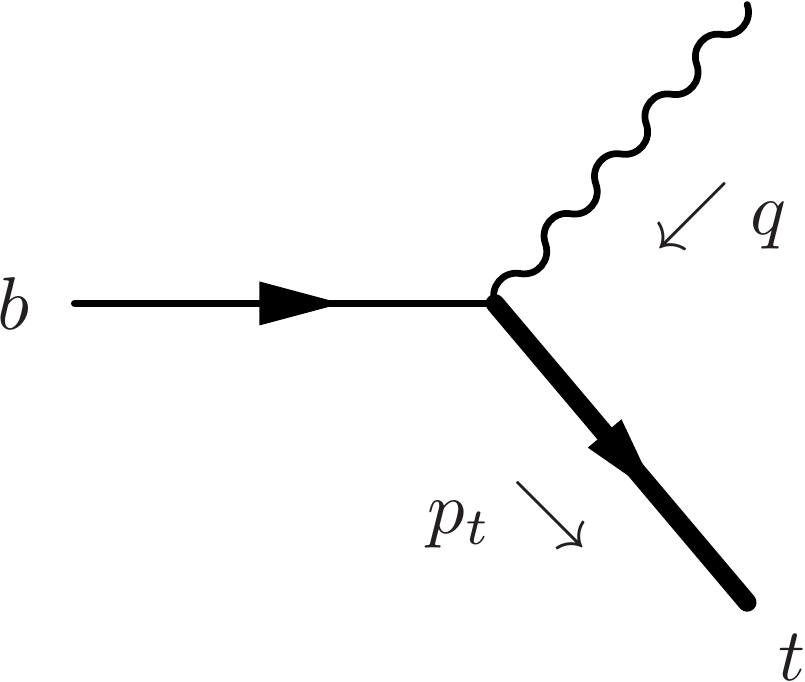}
  \caption{LO diagram for single top quark production. We show only the part
    relevant for corrections associated with the heavy quark line. The thick solid line
    denotes the top quark. The wavy line denotes 
    the off-shell $W$ boson that couples to the light quark line.\label{fig:5}}
  \end{figure}
Having described decay of the top quark in the previous subsection, we turn now to 
QCD corrections associated with its production.  In this subsection we treat the heavy 
quark line in the production process.  In the structure-function approximation in which we 
work, the heavy quark part of LO process is represented by the diagram in
Fig.~\ref{fig:5}.  The light quark part is omitted in this step; it is treated in the next 
subsection.  The light quark part (the upper vertex of Fig.~\ref{fig:born} )
can be thought effectively as the DIS ``leptonic'' part that is invisible to
the QCD corrections in the heavy quark line.  The process now resembles
charm-quark production in deep-inelastic neutrino scattering, for
which NNLO QCD corrections were calculated in
Ref.~\cite{1601.05430}.  We note that the process in
Fig.~\ref{fig:5} is related by crossing to the top quark decay process we discuss in the
previous section.  Many of the ingredients in the last section can be used here.

Following the previous section, we
define a resolution variable to isolate the unresolved part.   As discussed 
in Ref.~\cite{1601.05430}, the appropriate resolution variable in this case is a fully inclusive
version of beam thrust~\cite{Stewart:2009yx} or N-jettiness~\cite{1004.2489}, 
\begin{align}
  \label{eq:16}
  \tau_h =  \frac{2 \, p_X \mcdot p_n }{ m^2_t - q^2} ,\qquad \text{  with  } p_n  =
\Big(\bar{n} \cdot (p_t - q)\Big)    \frac{ n^\mu }{2} \,. 
\end{align}
It differs from the standard beam thrust or N-jettiness in that no
partition is imposed in the phase space of final-state radiation, as there is
only one collinear direction in the problem.  
This collinear direction is the beam (proton) direction associated with the 
bottom quark which enters the $Wtb$ vertex.
In Eq.~\eqref{eq:16},
$p_X$ is the momentum of total QCD radiation in the final state, and $p_n$
is a momentum aligned with the incoming beam whose large lightcone
component equals the large lightcone component of the incoming momentum
entering the $Wtb$ vertex.  Here the lightcone direction $n$ is chosen as the
direction of the incoming beam, and $\bar{n} = (1, -\vec{n})$, not to
be confused with the jet direction used in the last section.
Given the definition for $\tau_h$, the differential cross section for
any infrared-safe observable $O$ can be separated into resolved and
unresolved parts,
\begin{align}
  \label{eq:17}
    \frac{\dd\sigma_h}{\dd O} = & \int^{\tau_{h,\rm cut}}_0 \dd \tau_h \, \frac{\dd^2 \sigma_h}{\dd O
  \, \dd \tau_h} + \int^{\tau_{h,\rm max}}_{\tau_{h,\rm cut}} \dd \tau_h \, \frac{\dd^2 \sigma_h}{\dd O
  \, \dd \tau_h} 
\nn\\
= & \left. \frac{\dd\sigma_h}{\dd O} \right|_{\rm unres.} +
    \left. \frac{\dd\sigma_h}{\dd O} \right|_{\rm res.} \,.  
\end{align}
We use the subscript ``$h$'' to denote that QCD corrections to the light quark line are neglected.

Similar to the case of top quark decay, we can write a factorization
formula for the unresolved contribution, up to power corrections of
the form $\tau_{h,\rm cut} \ln^k \tau_{h,\rm cut}$,
\begin{align}
  \label{eq:18}
  \left. \frac{\dd\sigma_h}{\dd O} \right|_{\rm unres.}  =  & \int \dd z
  \,  \frac{\dd\sigma_h^\zero (z)}{\dd O} H_h(y,\mu) \int^{\tau_{h,\rm
  cut}}_0 \dd \tau_h \, \dd t \, \dd k_s   \, B_q( t, z,
  \mu) S_h(k_s, \mu) 
\nn\\
& \cdot \delta \left(\tau_h - \frac{t +2 k_s E_b}{m_t^2 - q^2} \right) + \cO( \tau_{h,\rm cut} \ln^k \tau_{h,\rm cut} )  \,,
\end{align}
where $E_b$ is the energy of the $b$ quark entering the $Wtb$ vertex. 
The derivation of this factorization formula is very similar to the
derivation of beam thrust in N-jettiness factorization. 
In Eq.~\eqref{eq:18}, $\dd\sigma_h^\zero(z)/\dd O$ is the Born level partonic
differential cross section for the process
\begin{align}
  \label{eq:19}
 b( zP_N)  +W^*( q )\to t(p_t)  \,,
\end{align} 
where $P_N$ is the momentum of the incoming hadron associated with the bottom quark.  
The definition of variable $y$ is $y = q^2/m_t^2 < 0$. The hard function for top
quark production can be related through analytic continuation in a straightforward 
way to the hard function for top quark decay, defined in Eq.~\eqref{eq:9},
\begin{align}
  \label{eq:20}
  H_h(y,\mu) = H_d(y + i 0, \mu) \,.
\end{align}

It is also possible to relate the heavy quark soft function to the 
decay soft function of Sec.~\ref{sec:qcd-corrections-top}.  They both involve a
timelike Wilson line and a lightlike Wilson line and a very similar
measurement function.  At one-loop
the heavy quark soft function can be calculated from the diagrams
\begin{align}
  \label{eq:21}
    S^\one_h(k_s,\mu) =\mu^{2 \e} \int \frac{\dd^{4 - 2 \e} k}{(2\pi)^{4 -  2
  \e}} (2\pi) \Theta(k^0) \delta(k^2) \delta( k_s - k \mcdot n) \left| \parbox[h]{0.18\textwidth}{
\includegraphics[width=0.15\textwidth]{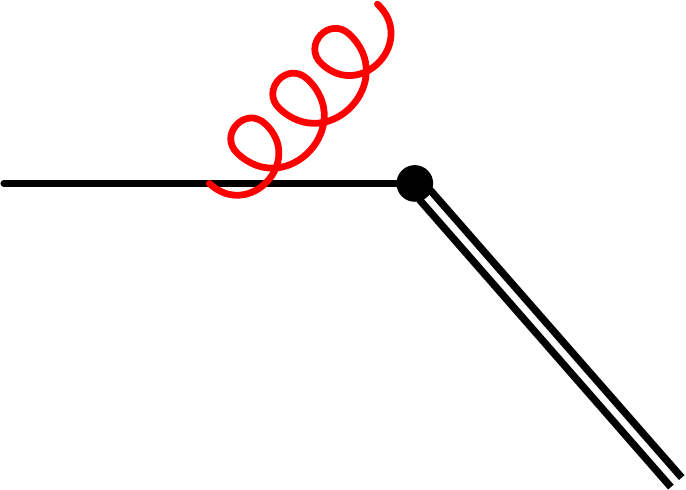} 
}
+ 
\parbox[h]{0.18\textwidth}{
\includegraphics[width=0.15\textwidth]{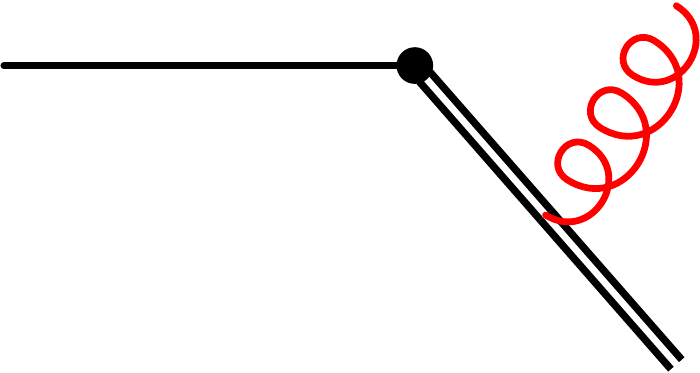} }
\right|^2 \,,
\end{align}
where the lightlike direction $n$ points in the incoming beam
direction.  Comparing with Eq.~\eqref{eq:10}, one may note that the timelike
Wilson line has been crossed from the initial state to the final state, whereas
the lightlike Wilson line from the final state to the initial
state.  This crossing leads only to a change of $\pm i
\varepsilon$ to $\mp i \varepsilon$ in the Feynman
prescription. The difference in the $i \varepsilon$ terms leads to a
sign difference in the Glauber phase $\exp(\pm i \pi f(\e) )$ for the
amplitudes, irrelevant at the cross section level. 

We must also deal with frame dependence of the soft function. The frame
dependence of the heavy quark soft function arises from the measurement
function, $\delta(k_s - k \mcdot n)$, just as for the decay soft
function.  Note that $k_s$ enters the observable through the
combination $2 E_J k_s$ for the top decay soft function, and $2 E_b k_s$
for the heavy quark soft function.  These combinations are Lorentz invariant,
as we may see from writing the measurement function for top decay as
\begin{align}
  \label{eq:22}
   2 E_J \delta( 2 E_J k_s - 2 k \mcdot p_b ) \,,
\end{align}
and for heavy quark production as
\begin{align}
  \label{eq:23}
  2 E_b \delta ( 2 E_b k_s - 2 k \mcdot p_b) \,.
\end{align}
Therefore, we can choose to define the heavy quark soft function in
the heavy quark rest frame, instead of the usual center of mass
frame. 
\begin{align}
  \label{eq:24}
  S_h(k_s , \mu) = S_d(k_s, \mu) \,, %\qquad \text{in the heavy quark rest frame}
\end{align}
in the heavy quark rest frame through all orders.  Moreover, 
\begin{align}
  \label{eq:25}
  E_b = \frac{m_t^2 - q^2}{2 m_t } 
\end{align}
in the heavy quark rest frame.  Now we can simply reuse the two-loop
soft function of Ref.~\cite{Becher:2005pd} for our heavy-quark-line
calculation.

The beam function is defined as the matrix element of a collinear field
in a hadron state~(proton in our case), with the virtuality $t = 2 p_n
\mcdot p_c$ of the measured beam jet~\cite{Stewart:2009yx}, where $p_c$ is the momentum of
final state collinear radiation,  and $p_n$ is defined in
Eq.~\eqref{eq:16}.  The beam function can be written as the convolution of
a perturbative coefficient function and the usual PDF,
\begin{align}
  \label{eq:27}
  B_i(t, x, \mu) = \sum_j \int \frac{\df\xi}{\xi} \, \cI_{ij} \left(t,
  \frac{x}{\xi}, \mu\right) f_j ( \xi, \mu) + \cO \left(
  \frac{\Lambda_{\rm QCD}^2}{t} \right) \,.
\end{align}
The one-loop quark-to-quark coefficient function can be calculated
through the diagrams
\begin{align}
  \label{eq:28}
    \cI_{qq}^\one\left(t, z, \mu\right) = & \int \frac{\dd^{4 - 2 \e} l}{(2\pi)^{4 -  2
  \e}} (2\pi) \Theta(l^0) \delta(l^2) \delta(t - 2 p_n \mcdot l )
  \delta \big( l \mcdot \bar{n}  - (1-z) p_n \mcdot \bar{n} \big)
\nn\\
&
\times
\left| \parbox[h]{0.28\textwidth}{
\includegraphics[width=0.25\textwidth]{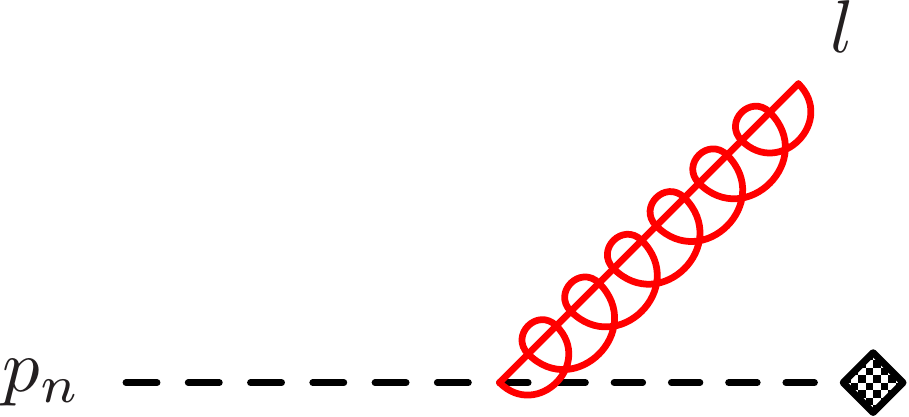} 
}
+ 
\parbox[h]{0.28\textwidth}{
\includegraphics[width=0.25\textwidth]{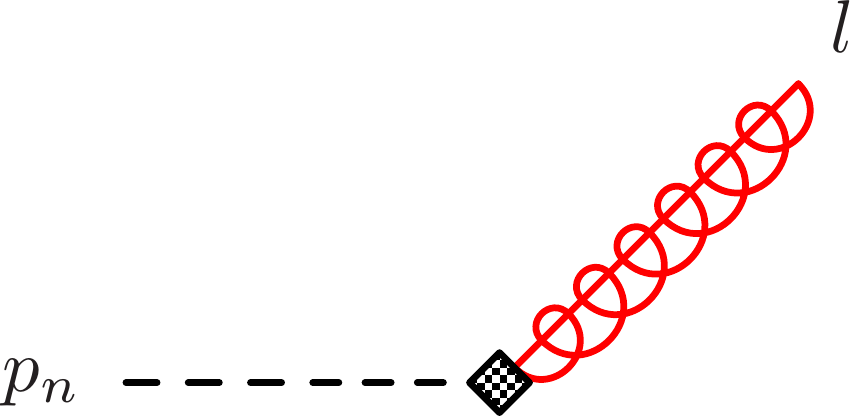} }
\right|^2 \,.
\end{align}
We also need the gluon-to-quark coefficient function at this order. The
quark beam function has been calculated through two loops~\cite{1401.5478}.  We quote
the result to one-loop here
\begin{align}
  \label{eq:29}
  \cI_{qq}(t, z, \mu) = &\delta(t) \delta(1-z) + \frac{\as}{2 \pi} C_F
  \left\{ 2\left[\frac{\ln (t/\mu^2)}{t} \right]_\star^{[t,\mu^2]} \delta(1-z) +
  \left[\frac{1}{t} \right]_\star^{[t,\mu^2]} \frac{(1+z^2)}{[1-z]_+}
\right.
\nn\\
&\left. 
+ \delta(t) \left[\frac{(1+z^2)}{[1-z]_+} - \frac{\pi^2}{6} \delta(1-z)
  + \left(1-z- \frac{1+z^2}{1-z} \ln z \right) \right] 
 \right\} + \cO(\as^2) \,,
\nn\\
\cI_{qg}(t,z,\mu) = & \frac{\as}{2 \pi} T_F \left\{ \left[\frac{1}{t}
  \right]_\star^{[t,\mu^2]} ( 1 - 2 z + 2 z^2) + \delta(t) \left[
                     (1 -2 z + 2 z^2) \left( \ln\frac{1-z}{z} - 1
                      \right) + 1 \right] \right\}  
\nn\\
& + \cO(\as^2) \,.
\end{align}
After substituting the expansion of hard, soft, and beam functions into the
factorization formula in Eq.~\eqref{eq:18}, one obtains the unresolved distribution to 
leading power in $\tau_h$.   Again, the dependence
on $\tau_h$ is very simple and can be integrated analytically. 

The calculation for the resolved contribution follows closely the
 decay calculation in Sec.~\ref{sec:qcd-corrections-top}. In fact, all the matrix
elements can be recycled from the last section. Again, we use dipole
subtraction to remove those sub-divergences which cannot be resolved by $\tau_h$.

\subsection{QCD corrections for production of a single top quark: light quark line}
\label{sec:qcd-corr-single-1}

For the QCD corrections associated with the light-quark line (the  
upper vertex of Fig.~\ref{fig:born} ),  we adopt the
method of ``Projection-to-Born'' in Ref.~\cite{1506.02660}.  The key
ingredients in this approach are the inclusive NNLO DIS coefficient
functions~\cite{vanNeerven:1991nn,Zijlstra:1992qd,Zijlstra:1992kj}, for
which a conveniently parametrized version is
available~\cite{vanNeerven:1999ca,vanNeerven:2000uj}.  The hadronic tensor
can be expressed in terms of three scalar form factors~\cite{1109.3717}
\begin{equation}\label{eq:p2b}
W_{\mu\nu}(x,Q^2)=(-g_{\mu\nu}+\frac{q_{\mu}q_{\nu}}{q^2})F_1(x,Q^2)
+\frac{\hat P_{\mu}\hat P_{\nu}}{P\cdot q}F_2(x,Q^2)+i\epsilon_{\mu\nu\alpha\beta}
\frac{P^{\alpha}q^{\beta}}{2P\cdot q}F_3(x,Q^2) \,.
\end{equation}  
Here $P$ is the momentum of the incident proton at the light-quark vertex,
$q$ is the momentum transfer carried by the virtual $W$ boson, 
$\epsilon_{\mu\nu\alpha\beta}$ is the completely antisymmetric tensor,
$Q^2=-q^2$, and Bjorken variable $x=Q^2/2P\cdot Q$.
The momentum $\hat P$ is defined as
\begin{equation}
\hat P_{\mu}=P_{\mu}-\frac{P\cdot q}{q^2}q_{\mu}\,.
\end{equation}
$F_i$ are structure functions for charged-current DIS which can be expressed
as convolutions of the parton distributions and the DIS coefficient functions.
In our case both $x$ and $q$ can be determined by kinematics at the
heavy-quark vertex.
By contracting this hadronic tensor with the squared matrix element for the
heavy-quark vertex, keeping phase space unintegrated for the top quark, we
can calculate the total cross sections and differential distributions of the
top quark. This procedure is similar to the NNLO calculations of Higgs boson
production via vector boson fusion in the double DIS approximation~\cite{1003.4451}.

\begin{figure}[ht]
  \centering
  \includegraphics[width=0.9\textwidth]{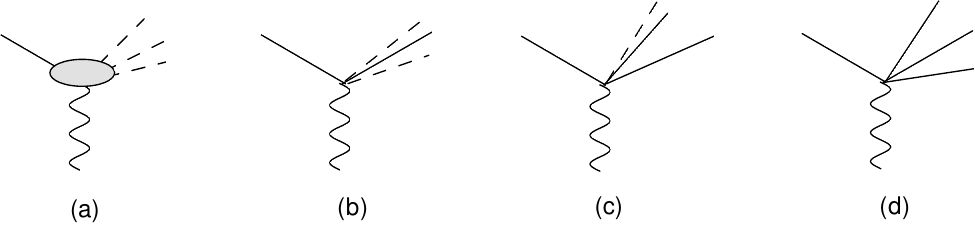}
  \caption{Schematic diagrams of the ``Projection-to-Born'' method, including
  the inclusive contributions, followed by separate contributions from the 
  double-unresolved region, the 
  single-unresolved region, and the fully-resolved region.}
  \label{fig:p2b}
\end{figure}

The method of ``Projection-to-Born'' was used later in Ref.~\cite{1506.02660}
to retain the jet activity at the light-quark vertex and applied to Higgs boson production. 
The spirit of the method is illustrated in Fig.~\ref{fig:p2b}.   
The full NNLO corrections can be separated into contributions from the 
double-unresolved region, single-unresolved region, and fully-resolved region
depending on the phase space of real radiation,
as sketched in the last three diagrams in Fig.~\ref{fig:p2b}.
Expressions for diagrams (c) and (d) can be
obtained from the NLO results for processes with 
one more hard radiation at Born level, similar to the phase space slicing method.
For the remaining contributions from the double-unresolved region, diagram (b),
expressions can be obtained by subtracting diagrams (c) and (d) from the inclusive
results represented by diagram (a).  Furthermore, since all radiation is 
unresolved there, the final state jet has Born-like kinematics and is
uniquely determined by $x$ and $q$.  
In practice, by a rearrangement of the different pieces, the final results
consist of two components. First is the NNLO structure function contribution
from Eq.~(\ref{eq:p2b}) with Born-like kinematics determined by $x$ and $q$.
Second is the
contribution from 2-jet production at NLO supplemented by a counter-term
contribution.  The counter-term is constructed in such a way that for
every event in the Monte Carlo integration of the 2-jet NLO piece,
a counter-event is generated with opposite weight and with Born-like
kinematics determined by $x$ and $q$.  The counter-events remove
contributions from the resolved region in the inclusive structure functions,
as well as make the NLO calculation numerically stable.    
For the real-virtual corrections needed for the 2-jet NLO calculation,
we extracted the one-loop helicity amplitudes
from DIS 2 jet production in Ref.~\cite{0904.2665}.

\subsection{Validation of the calculation}
\label{sec:valid-calc}
\begin{figure}[ht]
  \centering
  \includegraphics[width=0.45\textwidth]{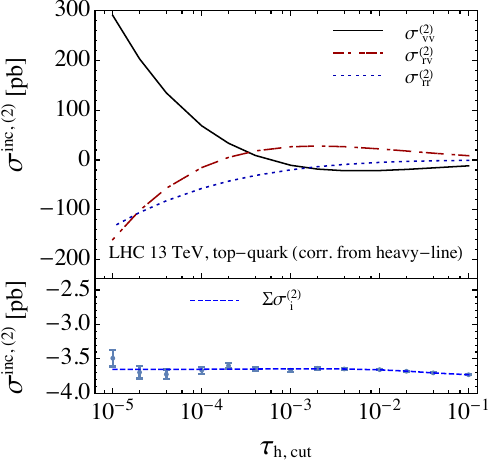}
  \caption{Various components of the NNLO corrections
  from the heavy-quark line for the total inclusive cross
  section as a function of the cut-off, for single top-quark
  production at 13 TeV.}
  \label{fig:dcheck1}
\end{figure}

\begin{figure}[ht]
  \centering
  \includegraphics[width=0.45\textwidth]{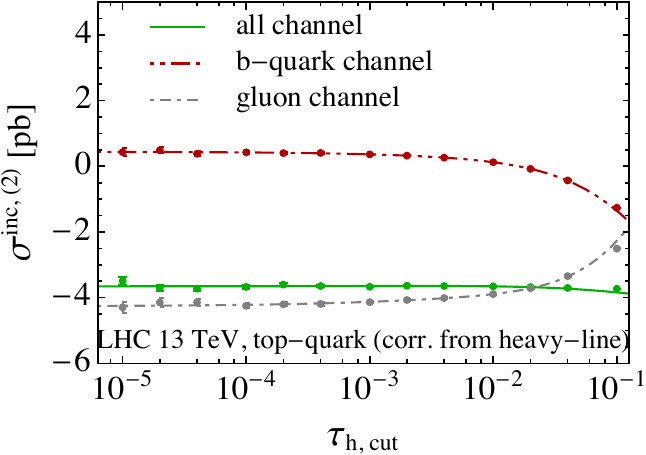}
  \caption{Cut-off dependence of different partonic channel 
  contributions at NNLO from the heavy-quark line for the total 
  inclusive cross section for top-quark production 
  at 13 TeV.}
  \label{fig:dcheck2}
\end{figure}

\begin{figure}[ht]
  \centering
  \includegraphics[width=0.458\textwidth]{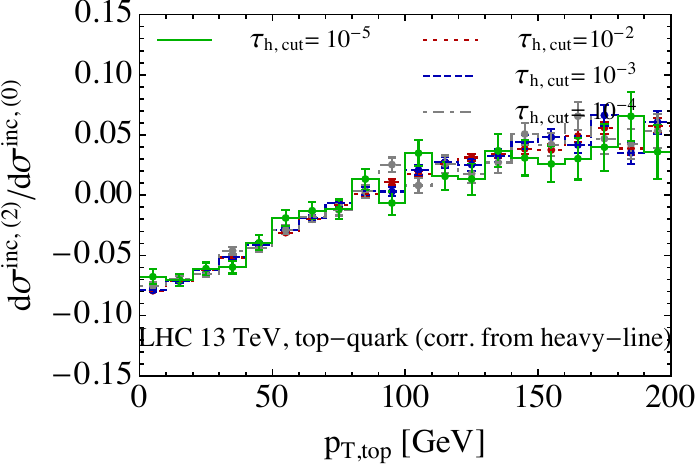}
  \hspace{0.4in}
  \includegraphics[width=0.442\textwidth]{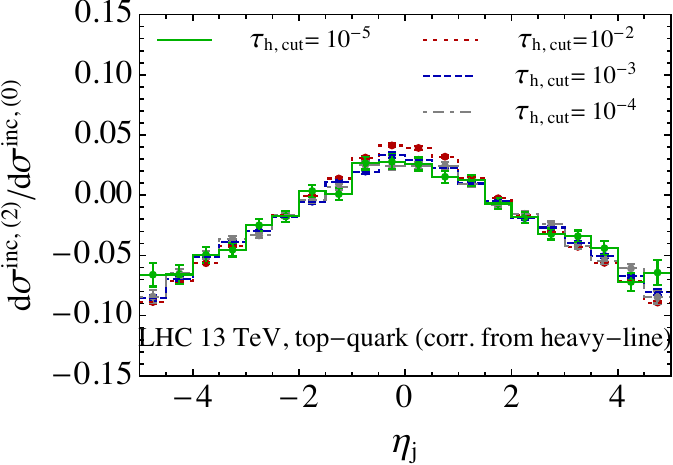}
  \caption{NNLO corrections from the heavy-quark line for 
  the transverse momentum distribution of the top quark (left),
  and the pseudorapidity distribution of the leading jet (right),
  for top-quark production at 13 TeV, with different choices 
  of the cut-off.}
  \label{fig:dcheck3}
\end{figure}

In the phase-space slicing method, we check the stability of 
various analytical expressions as well as the numerical implementations 
under variation of the small cut-off parameter.   
We should expect the results to converge smoothly to
the true NNLO corrections after large cancellations of individual pieces.
We demonstrate the cancellations for the heavy-quark line in Fig.~\ref{fig:dcheck1} 
for the case of NNLO corrections to the total inclusive cross sections.  In the upper
panel we show three contributions to the NNLO corrections: 
the below-cut-off unresolved contribution $\sigma_{VV}^{(2)}$,
the real-virtual part of the resolved contribution
$\sigma_{RV}^{(2)}$, and double-real parts of the resolved
contribution $\sigma_{RR}^{(2)}$.   The individual contributions vary considerably with
$\tau_{h,\rm cut}$, but the total contribution, shown in the lower panel, is 
stable and converges to a stable NNLO correction when
$\tau_{h,\rm cut}$ is small.  The cancellation of the three pieces is about one part 
out of a hundred.  One may notice that the power corrections are rather small 
even for $\tau_{h,\rm cut}$ as large as $10^{-1}$.  This small overall power correction results from an 
accidental cancellation of the power corrections from different partonic
channels at large $\tau_{h,\rm cut}$, as shown in Fig.~\ref{fig:dcheck2}.  Results for the $b$-quark
and the gluon channel show considerable dependence on the cutoff when $\tau_{h,\rm cut}\sim 10^{-1}$.
However these power corrections have a different sign and cancel largely in
the sum.  In Fig.~\ref{fig:dcheck3} we examine the dependence on
the cut-off in the differential distributions without decay.  We show the
transverse momentum of the stable top quark and the pseudorapidity of the
leading jet. The corrections are normalized to the LO distributions.
The error bars represent the estimated statistical 
uncertainties from the numerical integration.  There is good
agreement of the results when the cut-off is small.   
In practice we find optimal values of 
$\tau_{h,\rm cut}$ at about $10^{-4}\sim10^{-3}$ where
the power corrections are negligible and numerical
integration stability is preserved.               

For the resolved parts of all three NNLO calculations, we have cross checked our 
implementations with Gosam~\cite{1404.7096} and Sherpa~\cite{0811.4622} and 
found full agreement. The code
for the calculation involving top-quark decay is based on our previous
one used for calculation of the differential
width~\cite{1210.2808}. An independent calculation based on a different
infrared subtraction method was performed in Ref.~\cite{1301.7133}, and
it confirms our results.
We also checked explicitly that if we do not apply any selection cut, the
NNLO corrections from decay do not change the total event rate in our numerical
calculation, as expected from Eq.~(\ref{eq:2}).  For implementation of
the structure functions needed for the calculation of light-quark line,
we have compared our results with APFEL~\cite{1310.1394} and found good agreement. 
To compare our results with those in Ref.~\cite{1404.7116} for the case of
a stable top (anti-)quark, we calculated the 
NNLO total inclusive cross sections at 8 TeV using the same choices of parameters.  
We found a difference of $\sim 1 \%$ on the NNLO cross sections.  
With a refined comparison through private communications, we traced the source of this 
discrepancy to NNLO contributions associated with the heavy-quark line,
with the $b$-quark initial state.
All other parts in the NNLO corrections and all parts of the NLO
contributions agree between 
the two results within numerical uncertainties. It has not been
possible to further pin down the differences. 
We leave this issue for possible future investigation.

\begin{figure}[ht]
  \centering
  \includegraphics[width=0.9\textwidth]{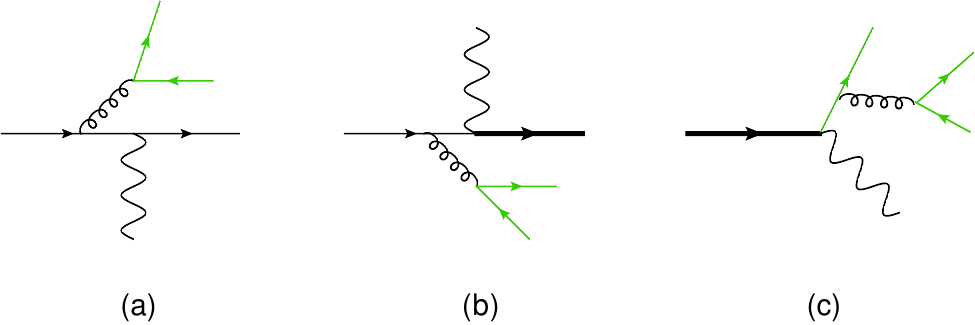}
  \caption{Feynman diagrams with a soft gluon splitting into a $b\bar b$
  pair, for NNLO corrections at the light-quark line, the heavy-quark line,
  and the top quark decay, respectively. The green fermion line represents
  a $b$ quark or anti-quark.}
  \label{fig:bjet}
\end{figure}

\begin{figure}[ht]
  \centering
  \includegraphics[width=0.5\textwidth]{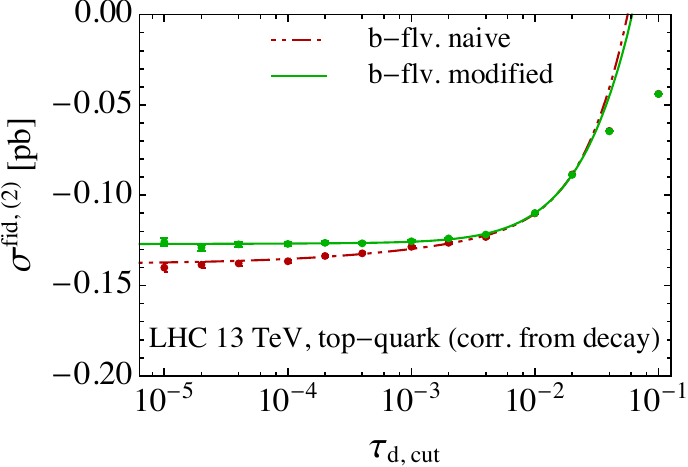}
  \caption{NNLO correction from top-quark decay 
  on the fiducial cross section as a function of the cut-off, for top-quark
  production at 13 TeV.  Error bars represent a scan over different
  cut-off values. The curves show fits to the points in the small
  cut-off region for two different flavor assignments. }
  \label{fig:ircheck}
\end{figure}

In calculations of the fiducial cross sections we also need a theoretically
well-defined flavor-jet algorithm for the $b$-quark jet. 
At the parton level, the definition of a $b$-quark jet has some level of ambiguity. 
Naively, the $b$-jet can be defined as a conventional jet whose total $b$-flavor
number is non-zero (counting the $b$ quark with $b$-flavor number $1$,
and $b$ anti-quark with $b$-flavor number $-1$). However, the resulting
jet cross section is not infrared safe in the zero-mass case.   In a
partonic configuration in which a soft gluon splits into a $b\bar{b}$
pair with large separation angle, the a $b$ quark may be clustered 
with other hard radiation and identified as a $b$-quark jet.
A modification of the $k_T$ algorithm to address the IR safety problem
of a $b$-quark jet is proposed in~\cite{0704.2999}. However, current
experimental measurements of single top-quark production at the LHC 
use the anti-$k_T$ algorithm~\cite{0802.1189}.   We do not adopt the
flavor-jet algorithm in~\cite{0704.2999}.  In our NNLO corrections,
the specific configurations which can lead to infrared safety issues appear in 
the diagrams shown in Fig.~\ref{fig:bjet}.   
In the first two diagrams, the splitting does not involve the primary $b$ quark
from decay of the top quark.  We simply assign a zero $b$-flavor number
for the bottom (anti-)quark in this splitting while keeping a non-zero
$b$-flavor number for the primary $b$ quark.  For the case of top quark decay
there can be two $b$ quarks and one $b$ anti-quark in the final state, as shown
by the third diagram.  In this case we first find a pair of $b$ quarks and an anti-quark, 
computing the invariant masses of the two $b {\bar b}$ pairs.   
Then we assign a zero $b$-flavor number for each $b/{\bar b}$ quark in the 
pair with the smaller invariant mass 
and a non-zero $b$-flavor number for the other $b$ quark.
After combining this modified flavor assignment with the anti-$k_T$ jet
algorithm, one can verify the infrared safety of our NNLO cross sections.  
An interesting test of infrared safety is made by checking the dependence of
the cross sections on the cut-off parameter.  In Fig.~\ref{fig:ircheck}
we show the NNLO corrections from top-quark decay on the fiducial cross section,
as a function of the cut-off $\tau_{d,\rm cut}$.  We can see the 
incomplete cancellation of cut-off dependence with the naive flavor
assignment, an indication of infrared problems of the algorithm.  In the 
modified case we observe convergence similar to that in Fig.~\ref{fig:dcheck2}.

\section{Cross sections and distributions for a stable top quark}
\label{sec:incl-cross-sect}
In this section we present predictions for the total inclusive cross sections and
differential cross sections for a top quark treated as an observable stable object.
%The calculations are done with a stable top quark since further decays do not
%affect the inclusive rates.  
The parameters used in our numerical calculations are listed here.  
We use a top quark mass of 172.5 GeV and a $W$ boson mass of
80.385 GeV.
We choose $|V_{tb}|=1$, $G_F=1.166379\times 10^{-5}{\rm GeV}^{-2}$
and the CT14 parton distribution functions (PDFs)~\cite{Dulat:2015mca}
with $\alpha_s(M_Z)=0.118$.  We specify below which perturbative order we 
use for the PDFs (LO, NLO, NNLO).  The nominal perturbative hard-scale choice 
is $\mu_R=\mu_F=m_t$ with
scale uncertainty evaluated by varying the two together over the range $0.5 < \mu/\mu_o < 2$.

\subsection{Total inclusive cross section}
\label{sec:incl-cross-sect-2}
\begin{figure}[h]
  \begin{center}
  \includegraphics[width=0.4\textwidth]{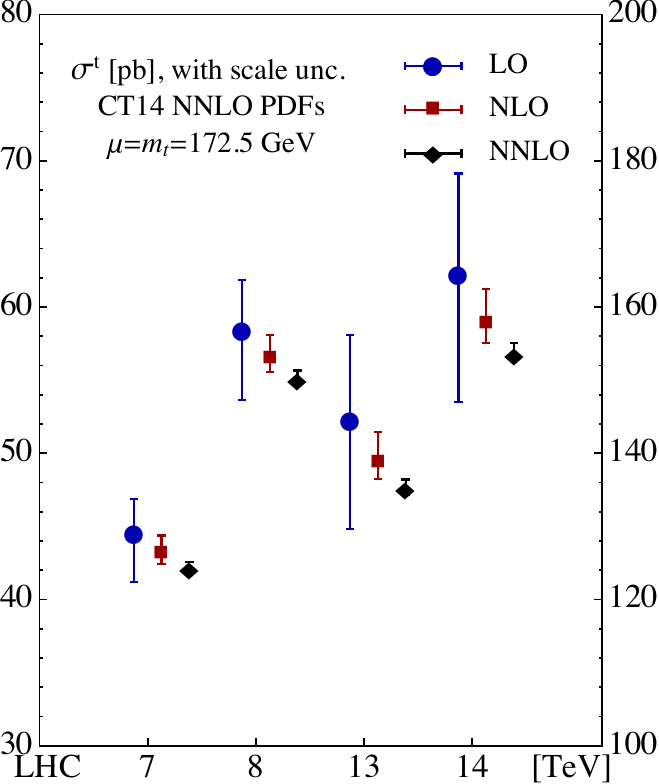}
  \hspace{0.4in}
  \includegraphics[width=0.4\textwidth]{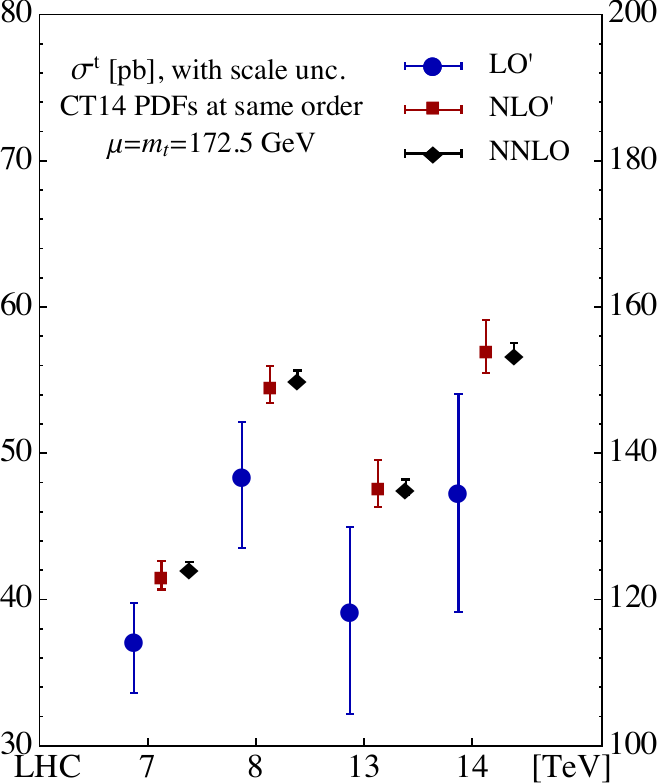}
  \end{center}
  \vspace{2ex}
  \caption{\label{fig:inc1}
   Inclusive cross sections for $t$-channel single top quark production at
   LO, NLO and NNLO with CT14 NNLO PDFs (left) and CT14 PDFs at same order (right),
   at the LHC with different center of mass energies.
   Error bars represent scale uncertainties obtained by varying the
   renormalization and factorization scale from $\mu_F=\mu_R=m_t/2$ to $2m_t$.}
\end{figure}

\begin{figure}[h]
  \begin{center}
  \includegraphics[width=0.4\textwidth]{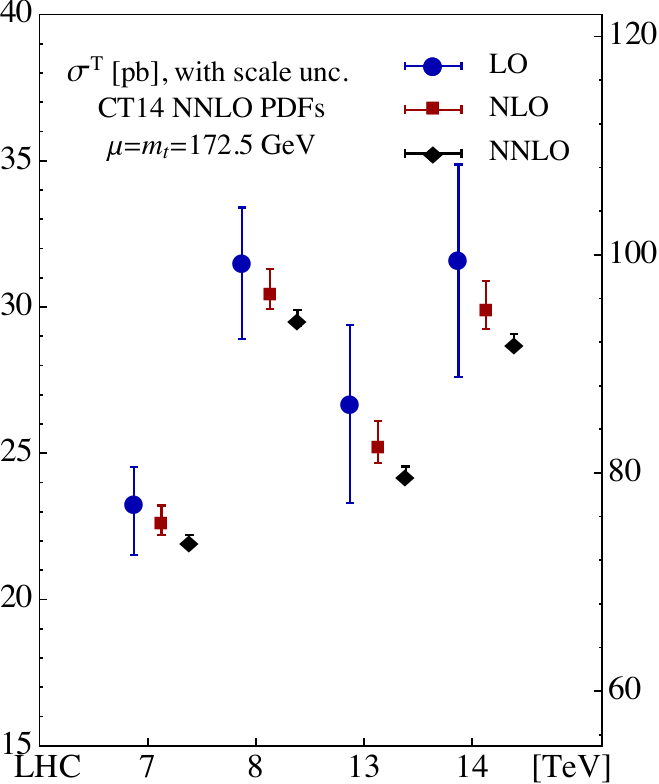}
  \hspace{0.4in}
  \includegraphics[width=0.4\textwidth]{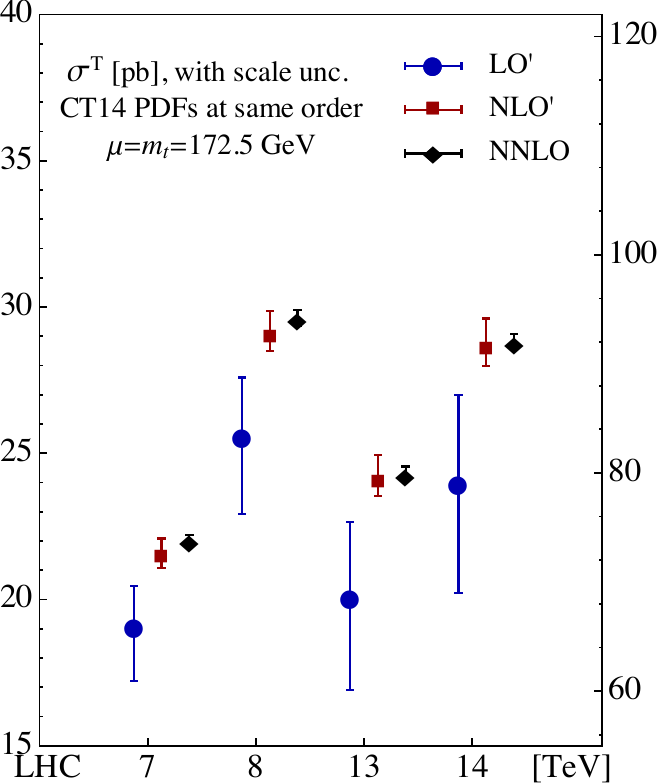}
  \end{center}
  \vspace{2ex}
  \caption{\label{fig:inc2}
   Inclusive cross sections for $t$-channel single top anti-quark production,
   similar to Fig.~\ref{fig:inc1}.}
\end{figure}

In Fig.~\ref{fig:inc1} we plot the total inclusive cross sections for single top-quark
production at the LHC with different energies.  For 7 and 8 TeV, the scale of  
cross sections is shown on the left-vertical axis; for 13 and 14 TeV the scale on the 
right-vertical axis is appropriate.
The predictions in the left side of Fig.~\ref{fig:inc1} are obtained with CT14 NNLO PDFs
throughout even though the hard matrix elements are computed at LO, NLO, and NNLO 
respectively.  The QCD corrections are negative when the same PDFs are 
used.  The NNLO corrections are about $2\sim 3$\% in general compared to
$3\sim 5$\% at NLO.   The error bars represent perturbative scale
variations at different orders.  Scale variations are reduced
by a factor of about 3 after the NNLO corrections are included. The remaining
uncertainties are generally at a level of one percent at NNLO,
e.g., +1.0\% and -0.6\% for top quark production at 13 TeV and
+1.1\% and -0.5\% for top anti-quark production.   
The predictions shown in the plot on the right side of Fig.~\ref{fig:inc1} are
obtained with CT14 PDFs at the associated orders, meaning with
LO PDFs for LO predictions and so on. In this case the LO predictions
drop significantly owing to the relatively smaller bottom-quark PDFs from
one-loop QCD evolution. The NNLO corrections are small possibly because
the process studied consists of similar components as for the deep
inelastic scattering (DIS) process used for determination of
PDFs. Thus it is expected that at least part of the QCD effects have
been absorbed into the fitting of PDFs. Figure~\ref{fig:inc2}
shows results for top anti-quark production at the LHC. Conclusions
concerning the size of QCD corrections and scale variations are similar to 
those for the top quark.  

\begin{figure}[h]
  \begin{center}
  \includegraphics[width=0.4\textwidth]{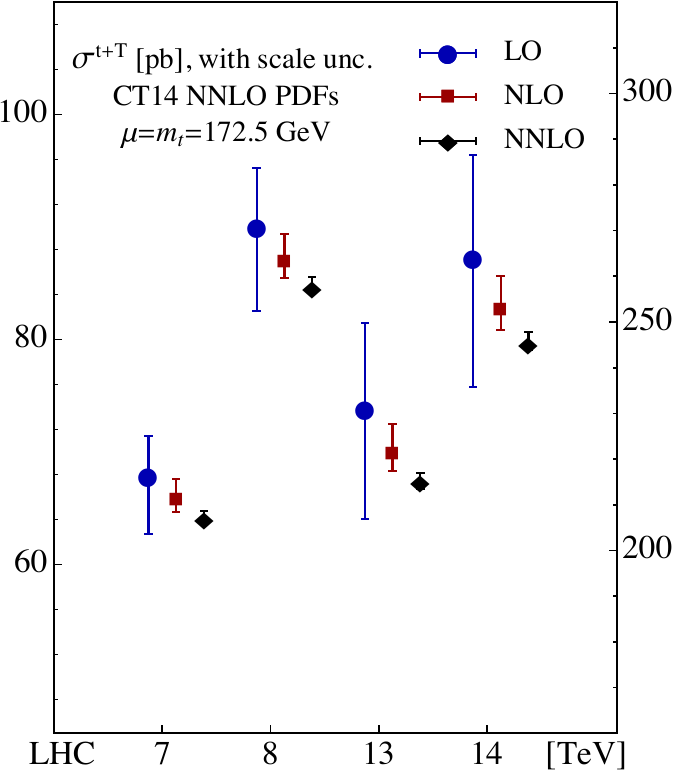}
  \hspace{0.4in}
  \includegraphics[width=0.4\textwidth]{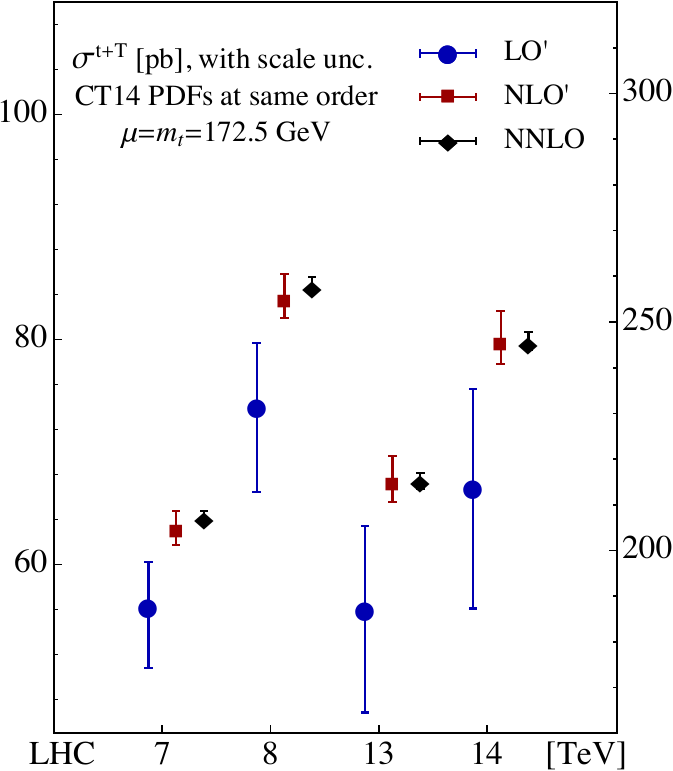}
  \end{center}
  \vspace{2ex}
  \caption{\label{fig:inc3}
   Sum of inclusive cross sections for $t$-channel single top quark and anti-quark
   production, similar to Fig.~\ref{fig:inc1}.}
\end{figure}

\begin{figure}[h!]
  \begin{center}
  \includegraphics[width=0.4\textwidth]{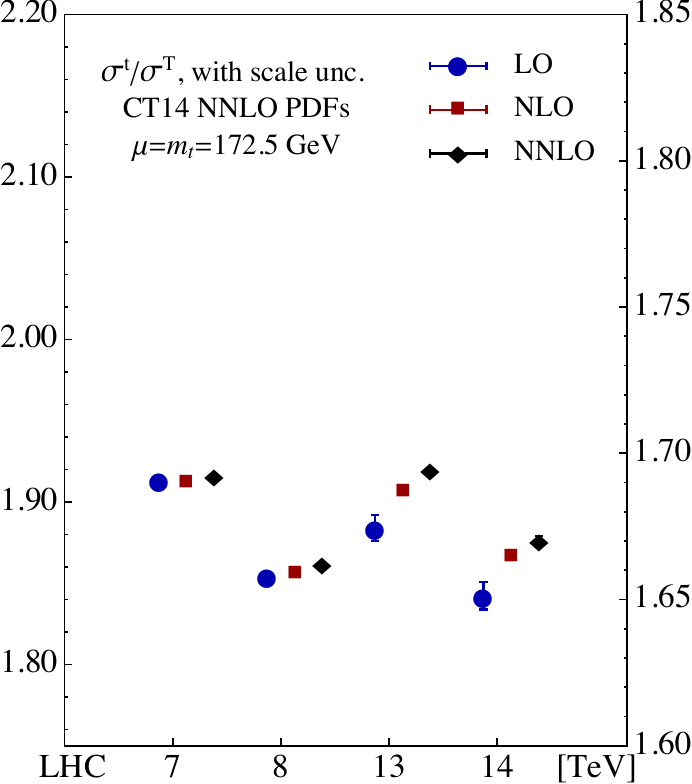}
  \hspace{0.4in}
  \includegraphics[width=0.4\textwidth]{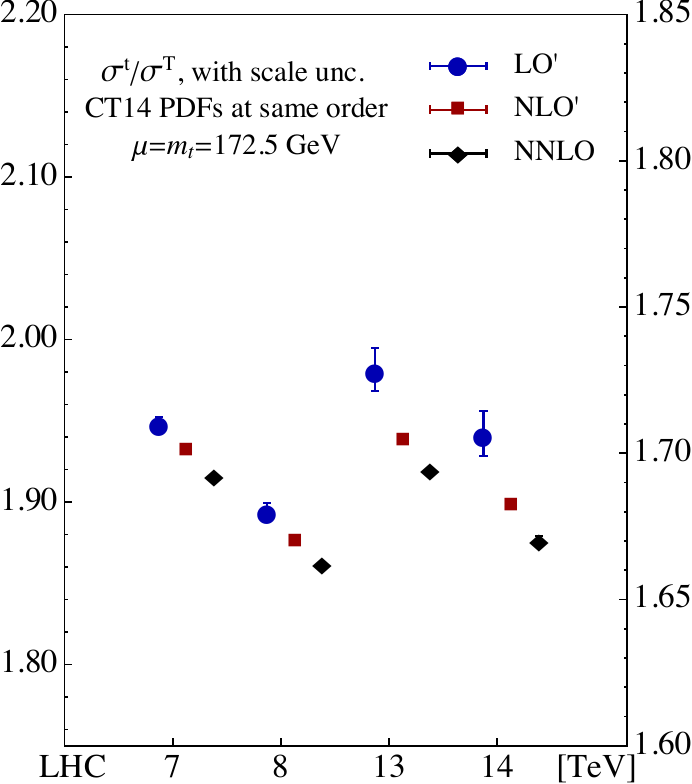}
  \end{center}
  \vspace{2ex}
  \caption{\label{fig:inc4}
   Ratio of inclusive cross sections for $t$-channel single top quark and anti-quark
   production, similar to Fig.~\ref{fig:inc1}.}
\end{figure}

We display the sum and ratio of the top quark and
anti-quark production cross sections in Fig.~\ref{fig:inc3} and Fig.~\ref{fig:inc4}, for
7 and 8 TeV with the scale on left-vertical axis, and 13 and 14 TeV with
the scale on right-vertical axis. In
both figures the scale variations are calculated by setting scales in top
quark and anti-quark production be the same and changing them simultaneously.
The behavior can be understood inasmuch as the QCD corrections in top quark and
anti-quark production are strongly correlated, as shown in Figs.~\ref{fig:inc1}
and~\ref{fig:inc2}.  The cross section ratios in Fig.~\ref{fig:inc4} are rather stable against 
QCD corrections. They change by at most 1\% from LO to NNLO if the same PDFs are used. 
The differences induced by PDFs at different orders are larger than the QCD corrections in 
general.  For completeness we provide numerical values of predictions with CT14 NNLO PDFs
in Table~\ref{tab:inclusive}. 

%%%%%%%%%%%%% begin figure 1 %%%%%%%%%%%%%%%%
%%%%%%%%%%%%% end   figure 1 %%%%%%%%%%%%%%%%

%%%%%%%%%%%%% begin figure 2 %%%%%%%%%%%%%%%%
%%%%%%%%%%%%% end   figure 2 %%%%%%%%%%%%%%%%

%%%%%%%%%%%%% begin figure 3 %%%%%%%%%%%%%%%%
%%%%%%%%%%%%% end   figure 3 %%%%%%%%%%%%%%%%

%%%%%%%%%%%%% begin figure 4 %%%%%%%%%%%%%%%%
%%%%%%%%%%%%% end   figure 4 %%%%%%%%%%%%%%%%
%
%
%%%%%%%%%%%%% begin table 1 %%%%%%%%%%%%%%%%
\begin{table}[h!]
\centering
\begin{tabular}{l|l|c|c|c} \hline
\multicolumn{2}{c|}{inclusive }  & LO & NLO & NNLO  \\  [1ex] 
\hline \hline 
\multirow{3}{*}{7 TeV} & $\sigma(t)\, {\rm [pb]}$ & $44.55_{-7.5\%}^{+5.3\%}$ &  $43.14_{-1.6\%}^{+2.9\%}$ & $42.05_{-0.6\%}^{+1.2\%}$   \\  [1ex] 
  & $\sigma(\bar t)\, {\rm [pb]}$ &  $23.29_{-7.6\%}^{+5.3\%}$ &  $22.57_{-1.5\%}^{+2.9\%}$ & $21.95_{-0.7\%}^{+1.2\%}$   \\  [1ex] 
  & $\sigma(t+\bar t)\, {\rm [pb]}$ &  $67.84_{-7.6\%}^{+5.3\%}$ &  $65.71_{-1.6\%}^{+2.9\%}$ & $64.00_{-0.6\%}^{+1.2\%}$   \\  [1ex] 
  & $\sigma(t)/\sigma(\bar t)$ &  $1.913_{-0.1\%}^{+0.1\%}$ &  $1.912_{-0.1\%}^{+0\%}$ & $1.916_{-0\%}^{+0.1\%}$   \\  [1ex] 
\hline
\multirow{3}{*}{8 TeV} & $\sigma(t)\, {\rm [pb]}$ & $58.41_{-8.1\%}^{+5.9\%}$ &  $56.46_{-1.6\%}^{+2.9\%}$ & $55.01_{-0.6\%}^{+1.2\%}$   \\  [1ex] 
  & $\sigma(\bar t)\, {\rm [pb]}$ & $31.52_{-8.3\%}^{+6.0\%}$ &  $30.41_{-1.6\%}^{+2.9\%}$ & $29.55_{-0.6\%}^{+1.2\%}$   \\  [1ex] 
  & $\sigma(t+\bar t)\, {\rm [pb]}$ &  $89.93_{-8.2\%}^{+5.9\%}$ &  $86.87_{-1.6\%}^{+2.9\%}$ & $84.57_{-0.6\%}^{+1.2\%}$   \\  [1ex] 
  & $\sigma(t)/\sigma(\bar t)$ &  $1.854_{-0.1\%}^{+0.1\%}$ &  $1.856_{-0.1\%}^{+0\%}$ & $1.861_{-0\%}^{+0.1\%}$   \\  [1ex]  
\hline
\multirow{3}{*}{13 TeV} & $\sigma(t)\, {\rm [pb]}$ & $144.5_{-10\%}^{+8.1\%}$ &  $138.8_{-1.7\%}^{+2.9\%}$ & $135.1_{-0.6\%}^{+1.0\%}$   \\  [1ex] 
  & $\sigma(\bar t)\, {\rm [pb]}$ & $86.34_{-10\%}^{+8.3\%}$ &  $82.28_{-1.6\%}^{+3.0\%}$ & $79.73_{-0.5\%}^{+1.1\%}$   \\  [1ex] 
  & $\sigma(t+\bar t)\, {\rm [pb]}$ &  $230.9_{-10\%}^{+8.2\%}$ &  $221.1_{-1.7\%}^{+3.0\%}$ & $214.8_{-0.6\%}^{+1.0\%}$   \\  [1ex] 
  & $\sigma(t)/\sigma(\bar t)$ &  $1.674_{-0.2\%}^{+0.3\%}$ &  $1.687_{-0.1\%}^{+0\%}$ & $1.694_{-0.1\%}^{+0\%}$   \\  [1ex] 
\hline
\multirow{3}{*}{14 TeV} & $\sigma(t)\, {\rm [pb]}$ & $164.4_{-10\%}^{+8.4\%}$ &  $157.8_{-1.7\%}^{+3.0\%}$ & $153.3_{-0.5\%}^{+1.1\%}$   \\  [1ex] 
  & $\sigma(\bar t)\, {\rm [pb]}$ & $99.60_{-11\%}^{+8.7\%}$ &  $94.77_{-1.6\%}^{+3.0\%}$ & $91.81_{-0.5\%}^{+1.0\%}$   \\  [1ex] 
  & $\sigma(t+\bar t)\, {\rm [pb]}$ &  $264.0_{-11\%}^{+8.5\%}$ &  $252.5_{-1.7\%}^{+3.0\%}$ & $245.1_{-0.5\%}^{+1.1\%}$   \\  [1ex] 
  & $\sigma(t)/\sigma(\bar t)$ &  $1.651_{-0.2\%}^{+0.3\%}$ &  $1.665_{-0.1\%}^{+0\%}$ & $1.670_{-0\%}^{+0.1\%}$   \\  [1ex] 
\hline
\end{tabular}
\caption{
Inclusive cross sections and their ratio for $t$-channel single top (anti-)quark production at
LO, NLO and NNLO with CT14 NNLO PDFs at the LHC with different center of mass energies.
Scale uncertainties are obtained by varying the renormalization and factorization scale from $\mu_F=\mu_R=m_t/2$ to $2m_t$.}
\label{tab:inclusive}
\end{table}
%%%%%%%%%%%%% end   table 1 %%%%%%%%%%%%%%%%

\begin{figure}[h!]
  \begin{center}
  \includegraphics[width=0.4\textwidth]{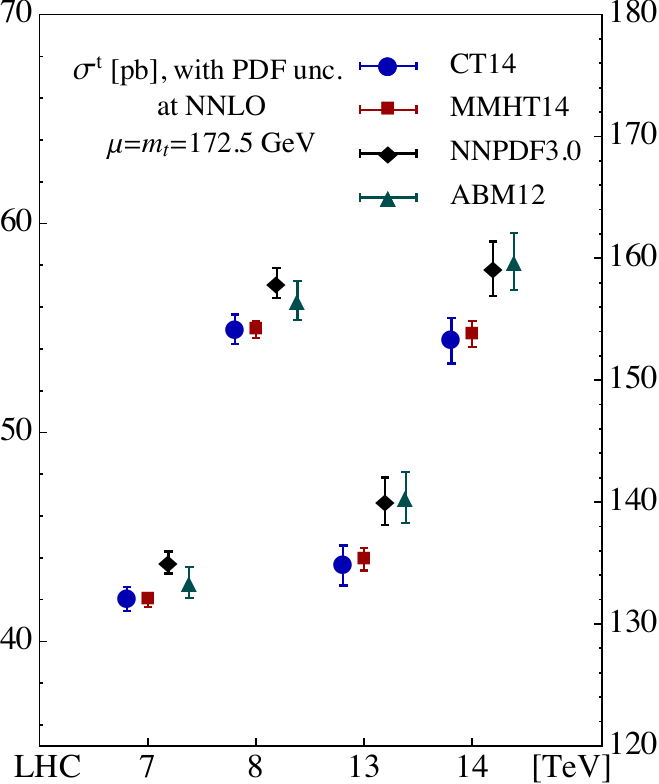}
  \hspace{0.4in}
  \includegraphics[width=0.4\textwidth]{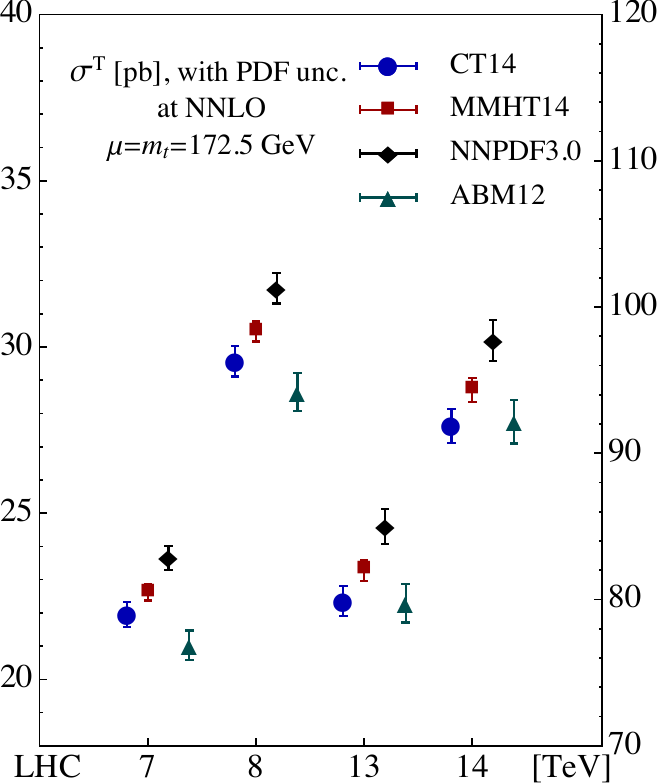}
  \end{center}
  \vspace{2ex}
  \caption{\label{fig:inc5}
   Inclusive cross sections for $t$-channel single top quark (left) and anti-quark (right) production at
   NNLO with various NNLO PDFs, for the LHC with different center of mass energies.
   Error bars represent $1\,\sigma$ PDF uncertainties.}
\end{figure}

\begin{figure}[h]
  \begin{center}
  \includegraphics[width=0.4\textwidth]{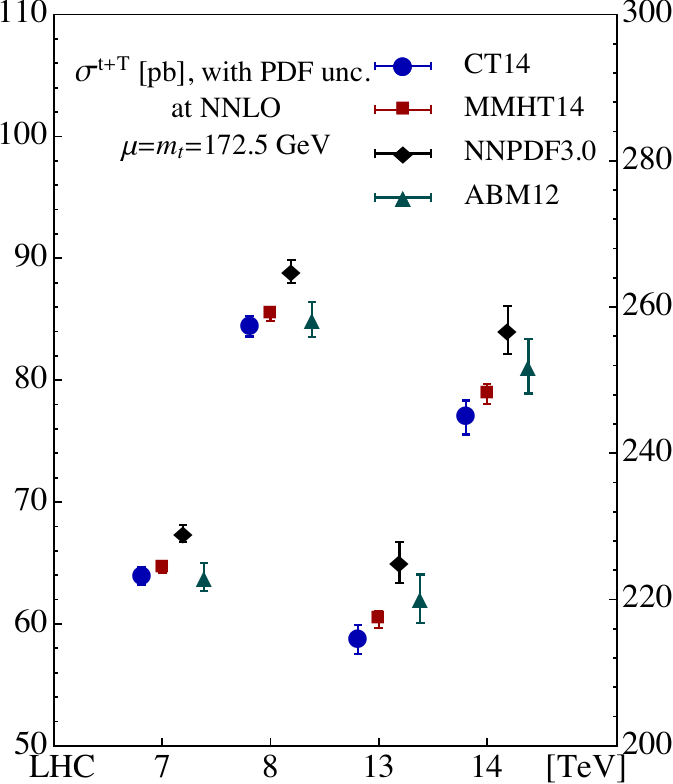}
  \hspace{0.4in}
  \includegraphics[width=0.41\textwidth]{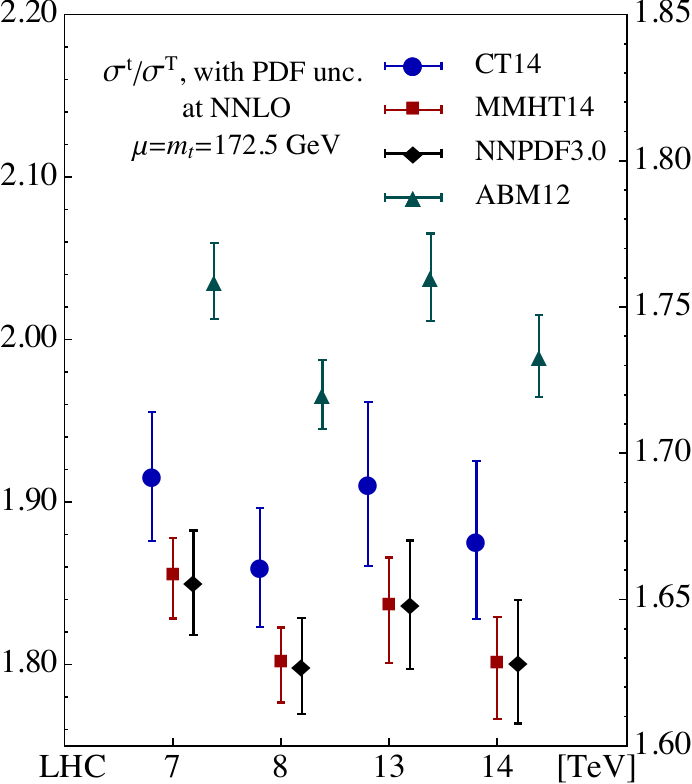}
  \end{center}
  \vspace{2ex}
  \caption{\label{fig:inc6}
   Sum (left) and ratio (right) of inclusive cross sections for $t$-channel single
   top quark and anti-quark production, similar to Fig.~\ref{fig:inc5}.}
\end{figure}

We show dependence of the total inclusive cross sections and
their ratios on different choices of PDFs in Figs.~\ref{fig:inc5} and~\ref{fig:inc6}, all
calculated at NNLO and with NNLO PDFs. The PDFs sets include CT14~\cite{Dulat:2015mca},
MMHT2014~\cite{Harland-Lang:2014zoa}, and NNPDF3.0~\cite{Ball:2014uwa},
all with $\alpha_s(M_Z)=0.118$, and ABM12~\cite{Alekhin:2013nda} with
the default $\alpha_s(M_Z)$ values. The error bars represent the $1\,\sigma$ PDF
uncertainties of individual groups.
The MMHT2014 results have the smallest PDF uncertainties among all groups.
The spread of predictions from different PDFs are especially large
for the top anti-quark production.  The spread can reach more than 10\%, as
shown by differences of the ABM12 and NNPDF3.0 predictions, 
amounting to deviations of about $3\sigma$, even if both error estimates are 
taken into account.
The discrepancies are even more pronounced in predictions of the
cross section ratios as shown in Fig.~\ref{fig:inc6}. The ABM12 PDFs
yield a much higher ratio compared to other three groups. Precise 
measurements of the cross section ratio from the LHC Run 2 can further
differentiate among these PDFs.

%%%%%%%%%%%%% begin figure 5 %%%%%%%%%%%%%%%%
%%%%%%%%%%%% end   figure 5 %%%%%%%%%%%%%%%%

%%%%%%%%%%%%% begin figure 6 %%%%%%%%%%%%%%%%
%%%%%%%%%%%%% end   figure 6 %%%%%%%%%%%%%%%%

\begin{figure}[h!]
  \begin{center}
  \includegraphics[width=0.4\textwidth]{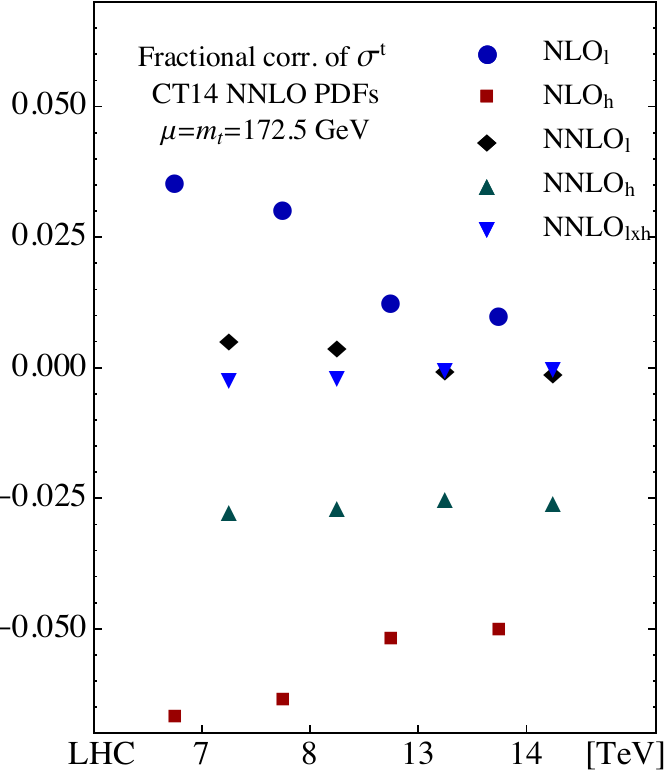}
  \hspace{0.4in}
  \includegraphics[width=0.4\textwidth]{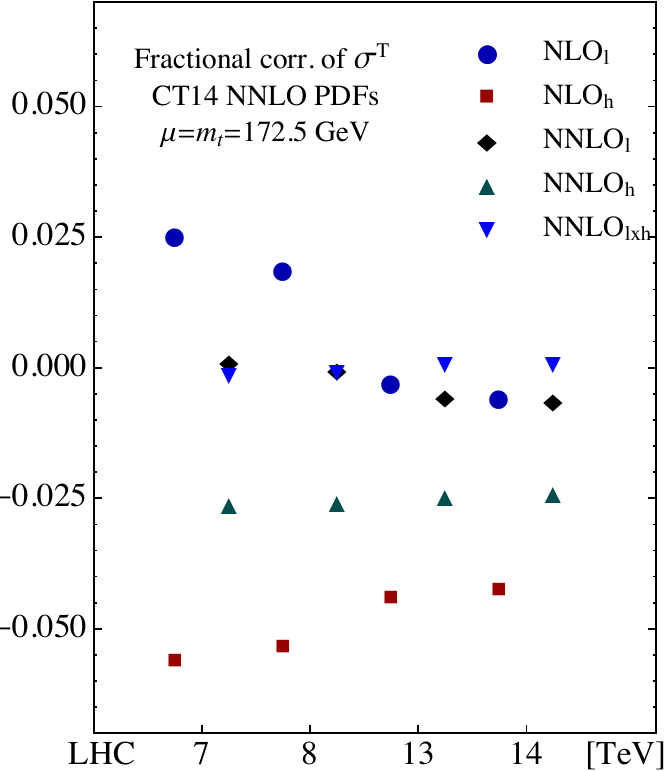}
  \end{center}
  \vspace{2ex}
  \caption{\label{fig:inc7}
   Fractional corrections for the inclusive cross sections of $t$-channel single top quark
   and anti-quark production at
   NLO and NNLO at the LHC with different center of mass energies, computed with CT14 NNLO PDFs,  
   and separated into component contributions.}
\end{figure}

In Fig.~\ref{fig:inc7} we show the fractional QCD corrections of
different gauge invariant pieces, including contributions from the light-quark
line, the heavy-quark line, and products of the two. The
latter starts at NNLO. We observe cancellations of QCD
corrections from the light and the heavy-line, driving the full corrections
to moderate negative values. The contributions from the heavy-quark
line dominate the NNLO corrections, while contributions from
the light-quark vertex and the products are almost negligible.  

%%%%%%%%%%%%% begin figure 7 %%%%%%%%%%%%%%%%
%%%%%%%%%%%% end   figure 7 %%%%%%%%%%%%%%%%

\subsection{Stable top quark differential cross sections}
\label{sec:diff-distr}

\begin{figure}[h!]
  \begin{center}
  \includegraphics[width=0.4\textwidth]{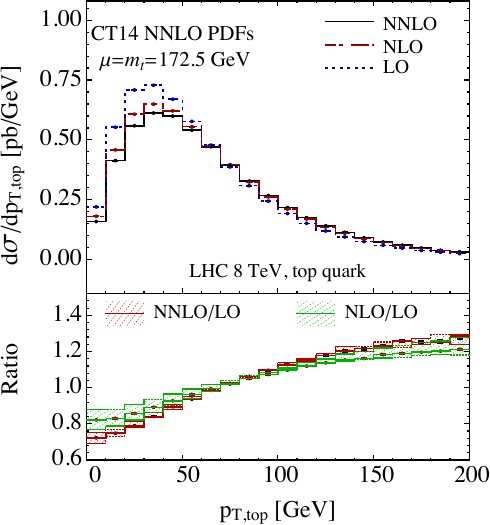}
  \hspace{0.4in}
  \includegraphics[width=0.4\textwidth]{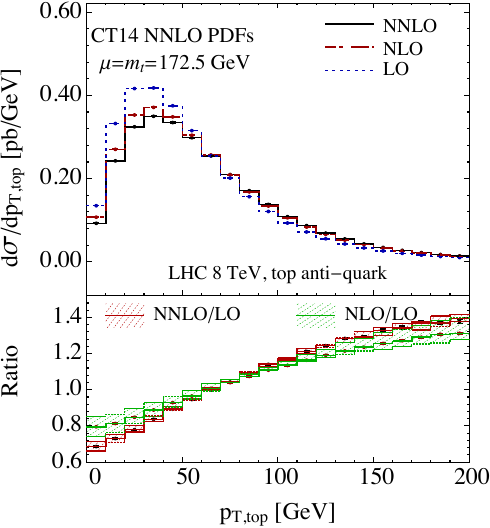}
  \includegraphics[width=0.4\textwidth]{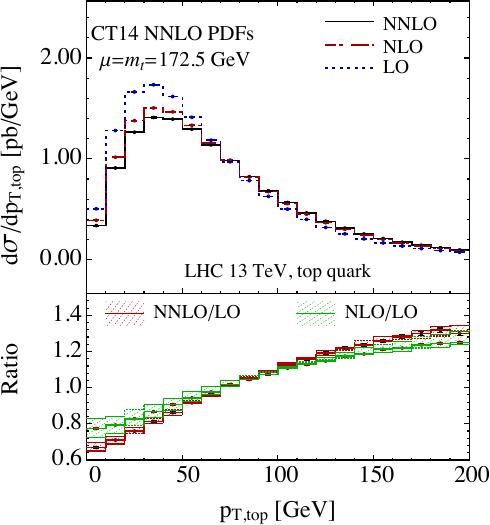}
  \hspace{0.4in}
  \includegraphics[width=0.4\textwidth]{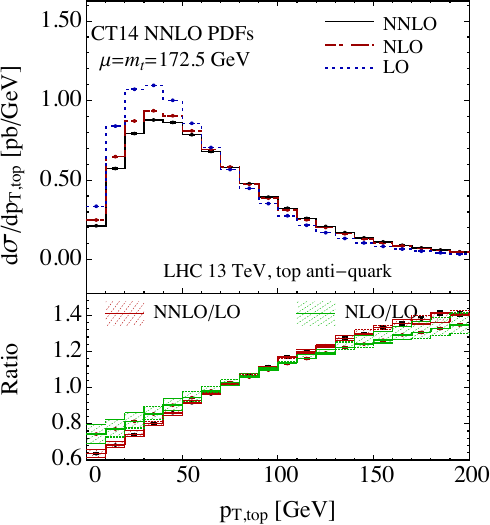}
  \end{center}
  \vspace{2ex}
  \caption{\label{fig:dis1}
   Predicted transverse momentum distribution of the top (anti-)quark
   from $t$-channel single top-quark production at 8 and 13 TeV.}
\end{figure}

We present transverse momentum distributions of the top quark
and anti-quark at 8 and 13 TeV in Fig.~\ref{fig:dis1}.
We show the distributions at various orders in
the upper panel, and the ratios of NLO and NNLO predictions to
the LO ones in the lower panel, with scale variations.  The QCD corrections 
can be negative or positive, depending on values of the transverse
momentum, and are smallest near 70 GeV. The corrections
are especially large in the regions of low and
high $p_{\rm T,top}$.  The NNLO corrections can be as
large as 10\%.  The scale variations are greatly reduced over 
the entire range of $p_{\rm T,top}$.   
%studied with the NNLO corrections.
The error bands from NLO and NNLO overlap over most of the region, 
suggesting that the scale variations provide a reasonable estimation
of the remaining perturbative uncertainties in this case. 
The QCD corrections are slightly larger for top anti-quark
production at high $p_{\rm T,top}$ compared to
top quark production. Dependence of the QCD corrections on
the center of mass energies is weak.  It may be argued that a 
$p_{\rm T,top}$ dependent dynamical scale such as $\sqrt {m^2_t + p^2_{\rm T,top}}$ 
is more appropriate for computations of the transverse momentum distribution, 
but we retain the central scale choice $m_t$ used elsewhere in this paper.  Differences 
from Fig.~\ref{fig:dis1} would be negligible at small $p_{\rm T,top}$ and more apparent 
in the region where $p_{\rm T,top} > m_t/2$.  
%similar to distributions shown later.

%%%%%%%%%%%%% begin figure 8 %%%%%%%%%%%%%%%%
%%%%%%%%%%%% end   figure 8 %%%%%%%%%%%%%%%%

\begin{figure}[h!]
  \begin{center}
  \includegraphics[width=0.4\textwidth]{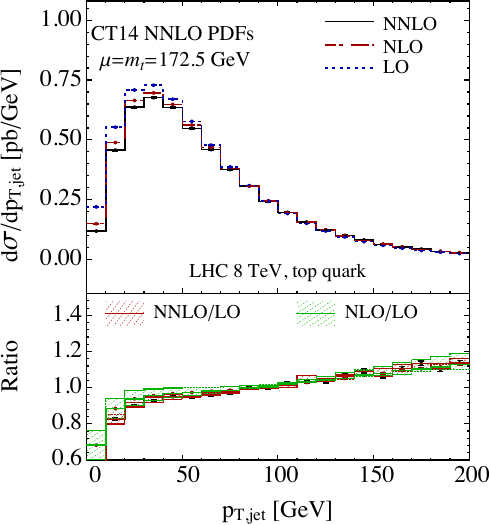}
  \hspace{0.4in}
  \includegraphics[width=0.4\textwidth]{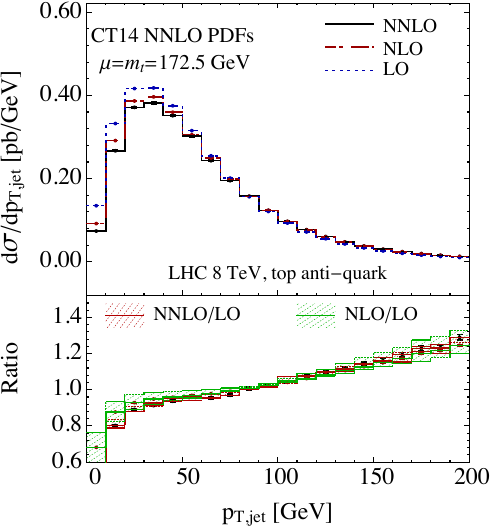}
  \includegraphics[width=0.4\textwidth]{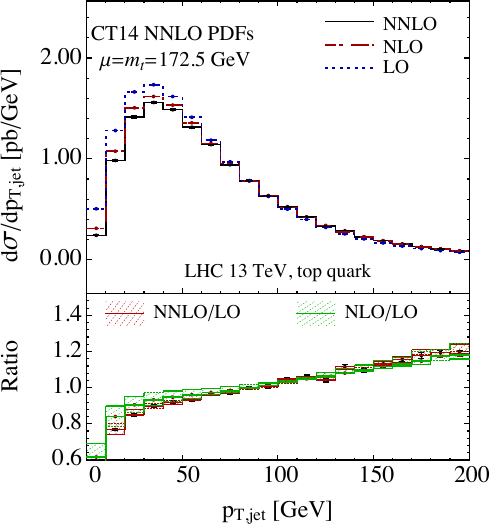}
  \hspace{0.4in}
  \includegraphics[width=0.4\textwidth]{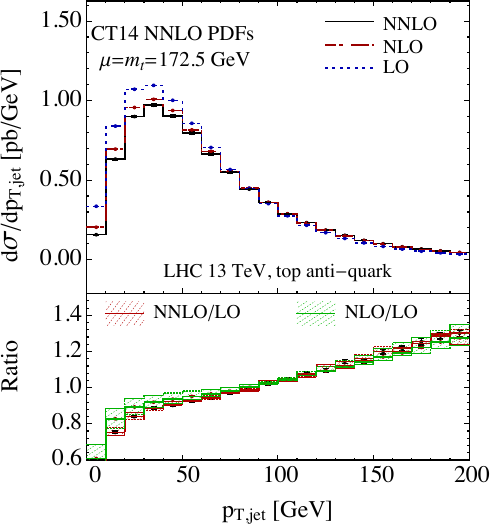}
  \end{center}
  \vspace{2ex}
  \caption{\label{fig:dis2}
   Predicted transverse momentum distribution of the leading-jet
   from $t$-channel single top-quark production at 8 and 13 TeV.}
\end{figure}

In Fig.~\ref{fig:dis2} we show the transverse momentum distributions
of the leading jet in top quark and anti-quark production.
We adopt the anti-$k_T$ jet algorithm with a
distance parameter of $D=0.5$.  At LO these are the same as those 
in Fig.~\ref{fig:dis1} since the top quark and jet are balanced in
transverse momentum at LO. The QCD corrections show a similar strong
shape dependence as the ones in Fig.~\ref{fig:dis1} but are smaller
in general.

\begin{figure}[h!]
  \begin{center}
  \includegraphics[width=0.4\textwidth]{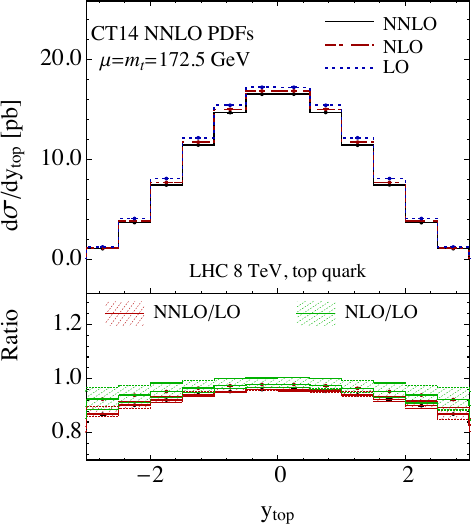}
  \hspace{0.4in}
  \includegraphics[width=0.4\textwidth]{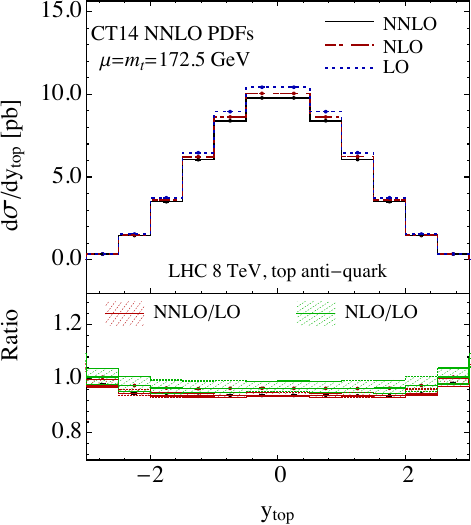}
  \includegraphics[width=0.4\textwidth]{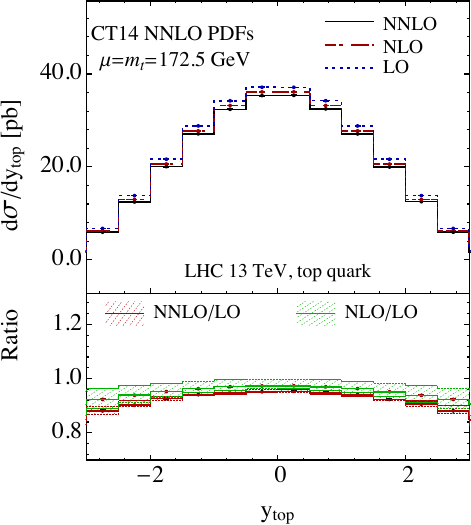}
  \hspace{0.4in}
  \includegraphics[width=0.4\textwidth]{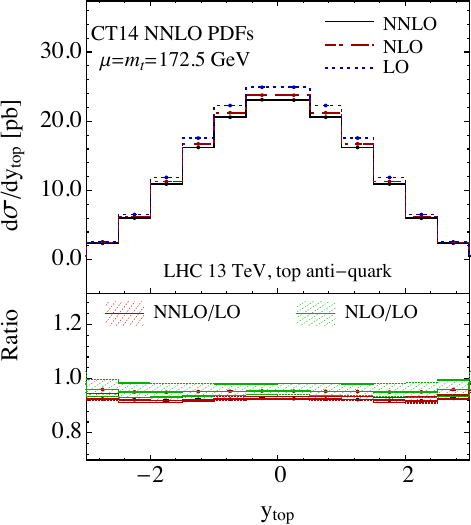}
  \end{center}
  \vspace{2ex}
  \caption{\label{fig:dis3}
   Predicted rapidity distribution of the top (anti-)quark
   from $t$-channel single top-quark production at 8 and 13 TeV.}
\end{figure}

\begin{figure}[h!]
  \begin{center}
  \includegraphics[width=0.4\textwidth]{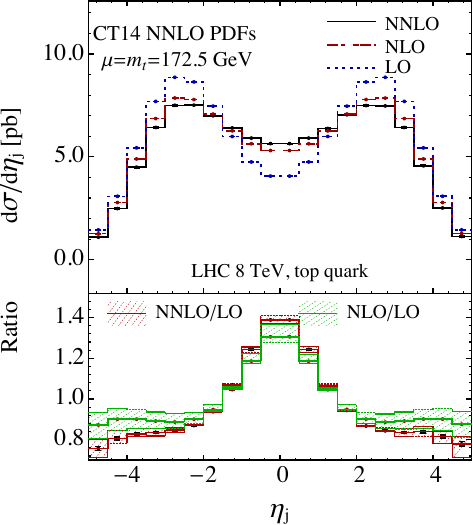}
  \hspace{0.4in}
  \includegraphics[width=0.39\textwidth]{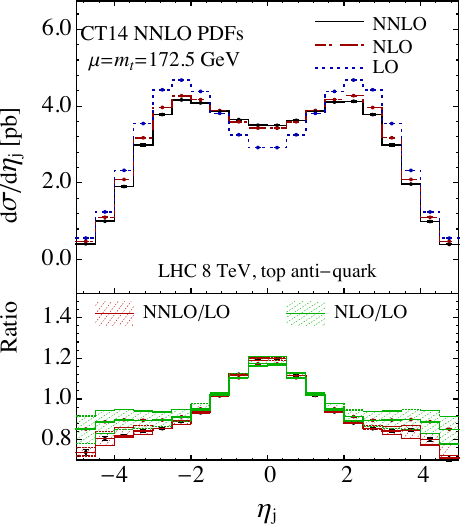}
  \includegraphics[width=0.4\textwidth]{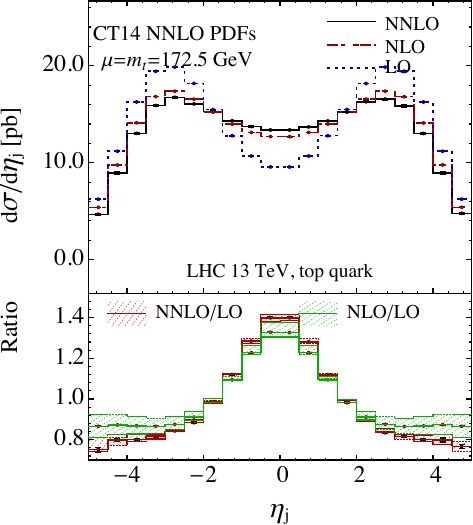}
  \hspace{0.4in}
  \includegraphics[width=0.4\textwidth]{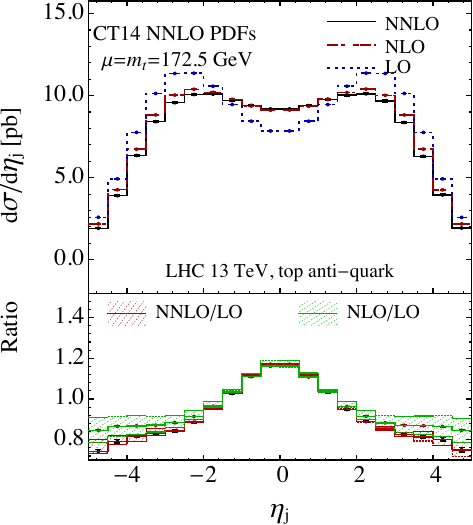}
  \end{center}
  \vspace{2ex}
  \caption{\label{fig:dis4}
   Predicted pseudorapidity distribution of the leading-jet
   from $t$-channel single top-quark production at the LHC 8 and 13 TeV.}
\end{figure}

Figures~\ref{fig:dis3} and~\ref{fig:dis4} show the rapidity
distributions of the top quark and anti-quark, and the pseudo-rapidity
distributions of the leading jet in top quark and anti-quark production,
respectively.  The QCD corrections have only a mild effect on 
the rapidity distributions.  The NNLO corrections
are moderate and at most 6\%.  The scale variations at NNLO are
almost contained within the NLO variation bands and are much smaller.
On the hand, the QCD corrections distort the shapes of the pseudo-rapidity 
distributions of the leading jet with respect to LO.  
They fill in the cross sections in the central region and
decrease them in the forward region.  The NNLO corrections
can be more than 10\%. In the central region of pseudo-rapidity
of the leading jet the QCD corrections are
more pronounced for a top quark than for an anti-quark. The scale 
variations are greatly reduced in all cases.

%%%%%%%%%%%%% begin figure 9 %%%%%%%%%%%%%%%%
%%%%%%%%%%%% end   figure 9 %%%%%%%%%%%%%%%%

%%%%%%%%%%%%% begin figure 10 %%%%%%%%%%%%%%%%
%%%%%%%%%%%% end   figure 10 %%%%%%%%%%%%%%%%

%%%%%%%%%%%%% begin figure 11 %%%%%%%%%%%%%%%%
%%%%%%%%%%%% end   figure 11 %%%%%%%%%%%%%%%%

\begin{figure}[h!]
  \begin{center}
  \includegraphics[width=0.4\textwidth]{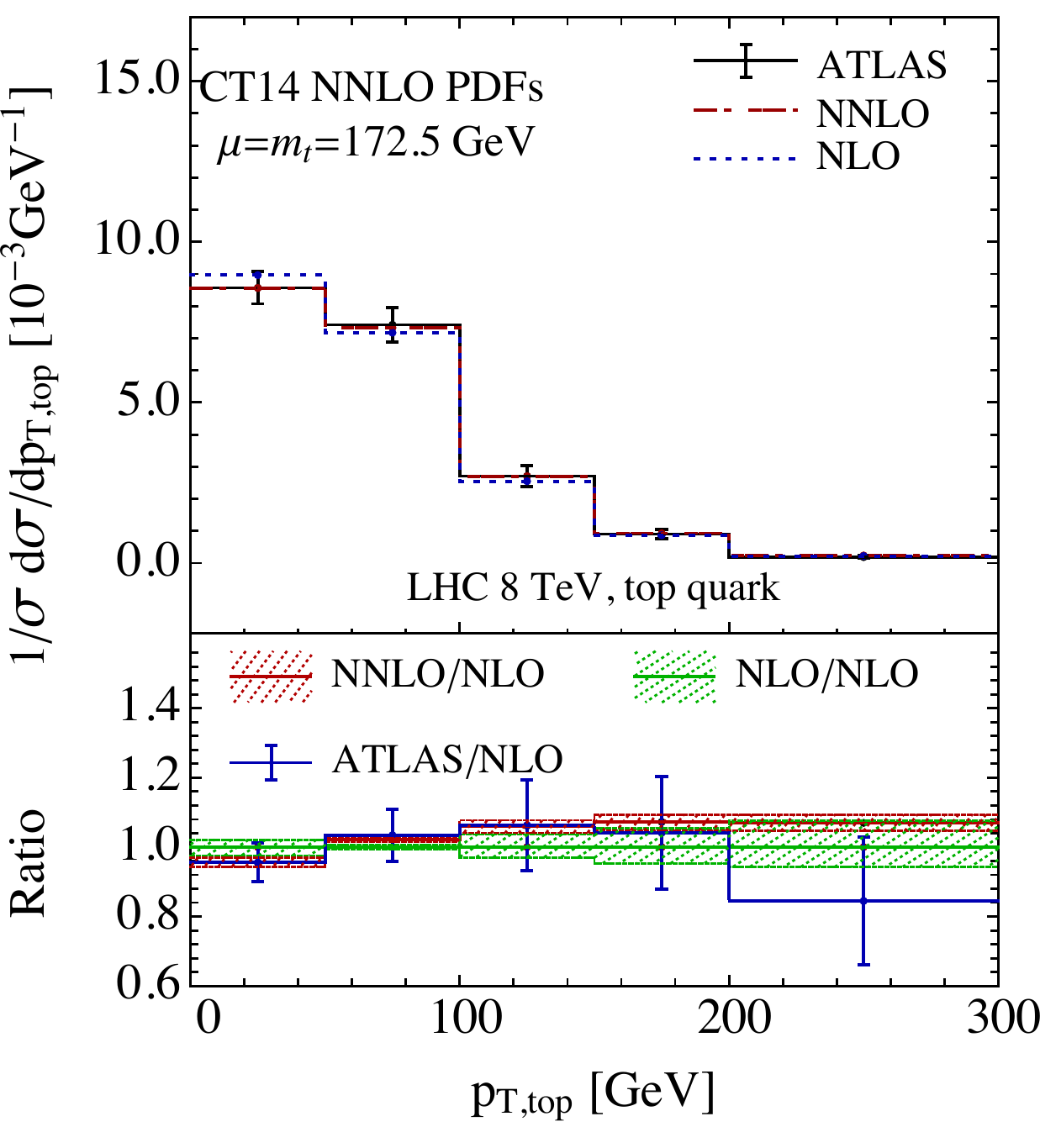}
  \hspace{0.4in}
  \includegraphics[width=0.4\textwidth]{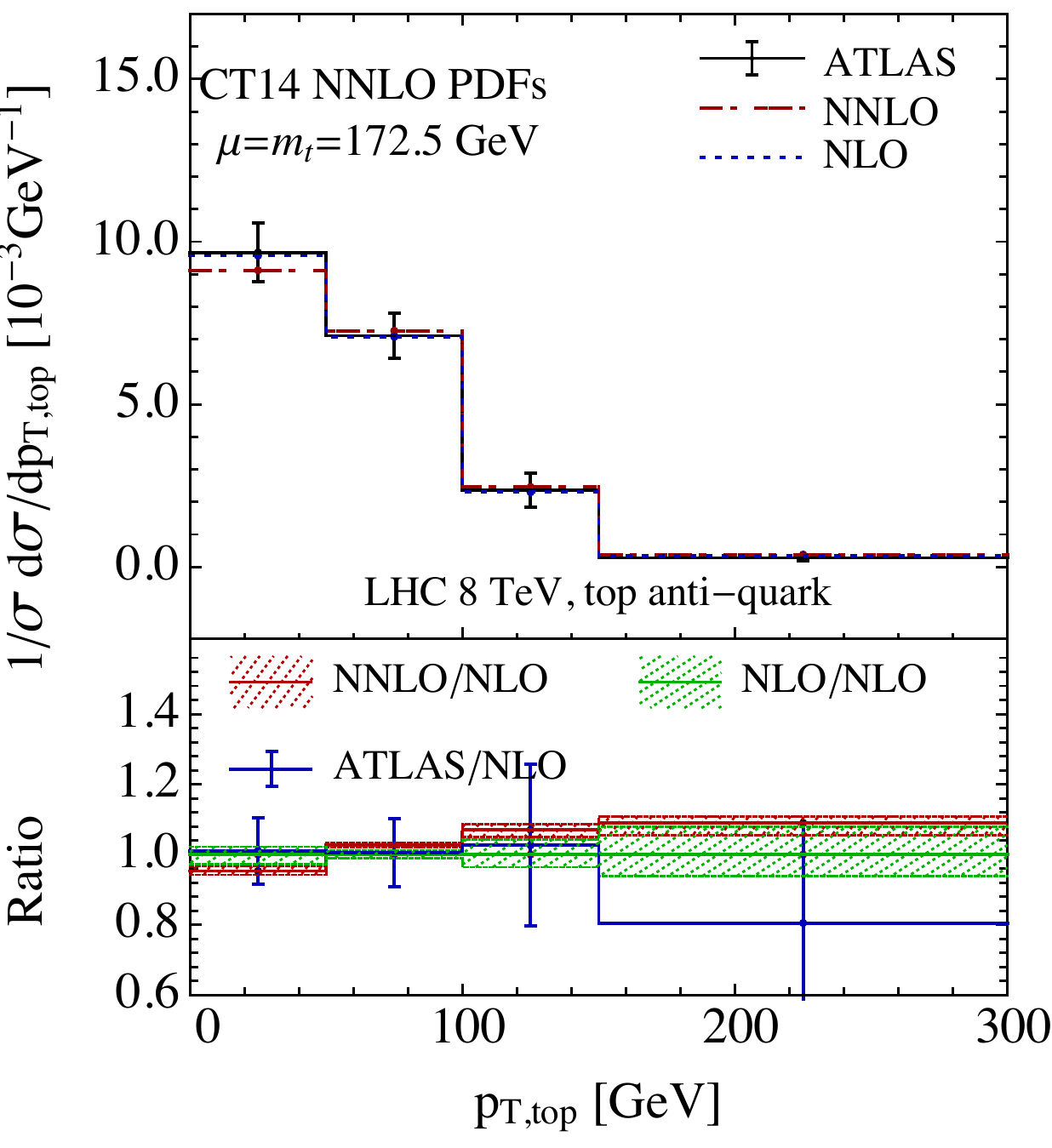}
  \end{center}
  \vspace{2ex}
  \caption{\label{fig:dis5}
   Predicted transverse momentum distribution of the top (anti-)quark
   from $t$-channel single top-quark production at 8 TeV compared 
   with the ATLAS data.}
\end{figure}

\begin{figure}[h!]
  \begin{center}
  \includegraphics[width=0.4\textwidth]{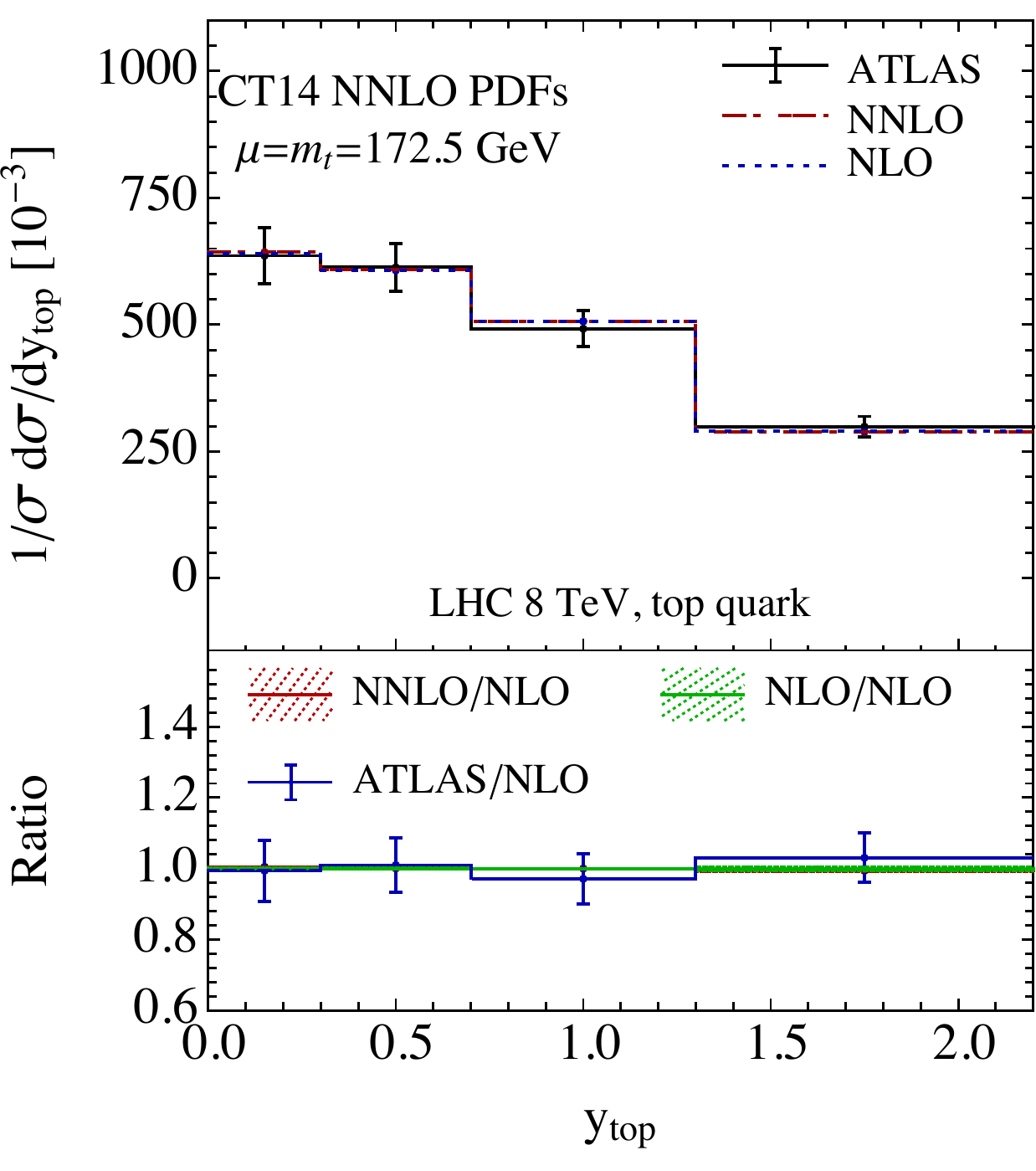}
  \hspace{0.4in}
  \includegraphics[width=0.4\textwidth]{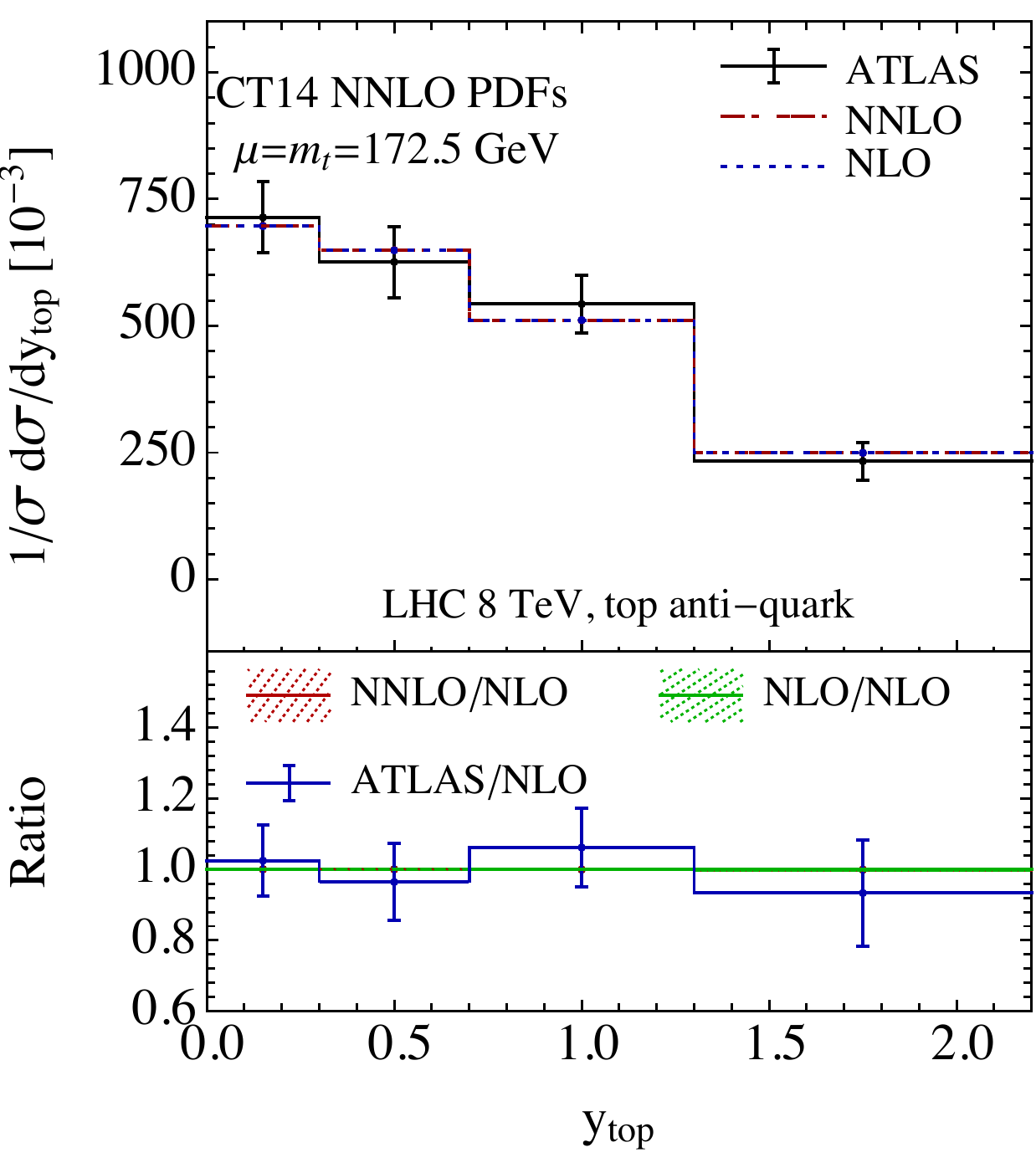}
  \end{center}
  \vspace{2ex}
  \caption{\label{fig:dis6}
   Predicted rapidity distribution of the top (anti-)quark
   from $t$-channel single top-quark production at 8 TeV compared   
   with the ATLAS data.}
\end{figure}

Measurements are available of parton-level differential distributions for 
$t$-channel single top-quark production at 8 TeV from the ATLAS
collaboration with a total luminosity of 20.2 ${\rm fb^{-1}}$~\cite{1702.02859}. We compare
our theoretical predictions based on CT14 NNLO PDFs with the ATLAS measurements
of the transverse momentum and rapidity distributions 
in Figs.~\ref{fig:dis5} and~\ref{fig:dis6}. 
We choose to compare with the normalized experimental
distributions for which the theoretical predictions are less sensitive to
the PDFs.  Ratios of predictions to the central values of the NLO prediction 
are shown in lower panels of all the plots.   The error bars in these plots
represent the total experimental uncertainties.  The hatched bands show the
scale variations.   The transverse momentum distributions presented in
Fig.~\ref{fig:dis5} indicate better agreement of the NNLO predictions with 
the central values of the ATLAS data compared to the NLO predictions.  
The last bin is the only exception, but the experimental uncertainties are large.  
Scale variations at NNLO are negligible compared to the experimental uncertainties.   
As noted earlier the QCD corrections have rather small effects on the rapidity 
distributions of the top (anti-)quark.  The lower panels of Fig,~\ref{fig:dis6} show 
good agreement of both the NLO and NNLO predictions with the ATLAS data.  

We should remark that the measured parton level distributions rely on an 
unfolding procedure, which depends on Monte Carlo (MC) event generators at NLO 
matched with parton showering~\cite{1207.5391}.  A more consistent comparison
of the NNLO predictions with the data should be made with unfolded
measurements based on the NNLO acceptance, even if the current
experimental uncertainties may already take into account part of
the bias introduced by the NLO unfolding procedure.      

%%%%%%%%%%%%% begin figure 12 %%%%%%%%%%%%%%%%
%%%%%%%%%%%% end   figure 12 %%%%%%%%%%%%%%%%

%%%%%%%%%%%%% begin figure 13 %%%%%%%%%%%%%%%%
%%%%%%%%%%%% end   figure 13 %%%%%%%%%%%%%%%%

\section{Fiducial cross section}
\label{sec:fiduc-cross-sect}
The fully differential nature of our calculation permits the computation of cross sections 
in a fiducial volume that matches closely the kinematic region of an experimental analysis.  
Such comparisons are potentially less subject to extrapolation uncertainties.
Experimental measurements of fiducial cross sections show a 
much weaker dependence on MC event generators and thus suffer less from
the related systematics.      
For calculations of the fiducial cross sections we assume the top quark
decays 100\% to $bW^+$ and set the $W$ boson leptonic decay branching
ratio to 0.1086 for one lepton family. We use a slightly different
top quark mass of 173.3 GeV here. 

We define the following fiducial phase space for the LHC at 13 TeV.
We use the anti-$k_T$ jet 
algorithm~\cite{0802.1189} with a distance parameter $D=0.5$.  Jets are required 
to have transverse momentum $p_T>40$ GeV and pseudorapidity $|\eta|<5$.  Following the CMS and ATLAS analyses, 
we require exactly two jets in the final state, meaning that events with additional jets are 
vetoed, and we require at least one of these to be a $b$-jet with $|\eta|<2.4$~\cite{CMS:2015jca}.
We demand the charged lepton to have a $p_T$ greater
than 30 GeV and rapidity $|\eta|<2.4$.  For the fiducial cross sections reported below we
include top-quark decay to only one family of leptons. Some of the numerical results
shown in this section are also reported in our earlier publication~\cite{Berger:2016oht}. 

\subsection{Total rate in the fiducial volume}
\label{sec:fiduc-cross-sect2}

%%%%%%%%%%%%% begin table 2 %%%%%%%%%%%%%%%%
\begin{table}[h!]
\centering
\begin{tabular}{l|l|c|c|c} \hline
\multicolumn{2}{c|}{fiducial [pb]}  & LO & NLO & NNLO  \\  [1ex] 
\hline \hline 
\multirow{3}{*}{$t$ quark} & total & $4.07_{-9.8\%}^{+7.6\%}$ &  $2.95_{-2.2\%}^{+4.1\%}$ & $2.70_{-0.7\%}^{+1.2\%}$   \\  [1ex] 
  & corr. in pro. &  & -0.79 & -0.24  \\ [1ex]
  & corr. in dec. &  & -0.33 & -0.13  \\ [1ex]
\hline
\multirow{3}{*}{$\bar t$ quark} & total & $2.45_{-10\%}^{+7.8\%}$ & $1.78_{-2.0\%}^{+3.9\%}$ & $1.62_{-0.8\%}^{+1.2\%}$  \\  [1ex] 
  & corr. in pro. &  & -0.46 & -0.15  \\ [1ex]
  & corr. in dec. &  & -0.21 & -0.08  \\ [1ex]
\hline
\end{tabular}
\caption{Fiducial cross sections for top (anti-)quark production with decay at 13 TeV 
at various orders in QCD with a central scale choice
of $m_t$ in both production and decay. The scale uncertainties correspond to
a quadratic sum of variations from scales in production and decay, and
are shown in percentages. Corrections from purely production and purely decay
are also shown.}
\label{tab:fiducial}
\end{table}
%%%%%%%%%%%%% end   table 2 %%%%%%%%%%%%%%%%

Table~\ref{tab:fiducial} shows our predictions of the fiducial cross sections
at different perturbative orders, with scale variations shown in
percentages.
We vary the renormalization and factorization scales $\mu_R=\mu_F$ in the top-quark 
production stage, and the renormalization scale in the decay stage, 
independently by a factor of two around the nominal scale choice.
The resulting scale variations are added in
quadrature to obtain the numbers shown in Table~\ref{tab:fiducial}. We also
show the QCD corrections from production and decay separately as defined in
Eq.~(\ref{eq:2}).  All results shown in Table~\ref{tab:fiducial} pertain to the
central scale choice $m_t$,  as for the inclusive cross sections.
The NNLO corrections from the product of $\mathcal{O}(\alpha_{\sss S})$
production and $\mathcal{O}(\alpha_{\sss S})$ decay can be derived by subtracting the 
above two contributions from the full NNLO corrections.

The NLO correction amounts to a decrease of the fiducial cross section by 
almost 30\% for top quark production.  A change this large requires investigation of the 
NNLO QCD corrections to examine convergence of the series.  The numbers in 
the Table indicate that the full NNLO correction drops the fiducial cross section 
by another 8.5\% relative to the NLO value.  
The corrections from decay are
half of the corrections from production in general. The NNLO corrections
from products of production and decay are similar in size to those from decay,  
but with different sign. The scale variations have been reduced
by a factor of about 3 to $\sim 1\%$ at NNLO. However, for fiducial cross sections,
the error bands from LO, NLO, and NNLO do not overlap each other
suggesting that scale variations underestimate the true 
perturbative uncertainties in this case.  The size of QCD corrections
are similar for top anti-quark production.   The ratio of fiducial cross sections for top quark 
and anti-quark production are 1.661, 1.657, and 1.667 at LO, NLO, and NNLO,
respectively. Therefore these charge ratio observables are stable against QCD
corrections even in the fiducial phase space.  

In experimental analyses, the total inclusive cross sections are usually 
determined through extrapolation of the fiducial cross sections based on
acceptance estimates obtained from MC simulations.  We can use the numbers
shown in Tables~\ref{tab:inclusive} and~\ref{tab:fiducial} to derive the
parton-level acceptance at various orders.  For top quark production, the
acceptances are 0.0283, 0.0214, and 0.0201 at LO, NLO, and NNLO respectively.
The NNLO corrections can change the acceptance by $6\%$ relative to the NLO value. 
This change also propagates into the measurement of the total
inclusive cross section through extrapolation.

A comment here is appropriate on the size of QCD corrections and the choice of the 
QCD hard scale.  
With fiducial cuts applied, the jet veto introduces another hard scattering
scale of $p_{T,veto}= 40$ GeV in addition to $m_t$.  A QCD scale choice 
$(p_{T,veto}m_t)^{1/2}\sim m_t/2$ may therefore be appropriate, especially at lower 
perturbative orders where the gluon splitting contributions are absorbed into the 
bottom-quark PDF.  Alternative results with a central scale choice of $m_t/2$ in 
production, with the central scale $m_t$ retained in decay, show  
better convergence of the series, although the NNLO predictions are almost unchanged.
It would be worthwhile to resum the logarithmic contributions related
to the scales $p_{T,veto}$ and $m_t$.

\subsection{Distributions within the fiducial region}
\label{sec:fiduc-cross-sect3}
\begin{figure}[h!]
  \begin{center}
  \includegraphics[width=0.4\textwidth]{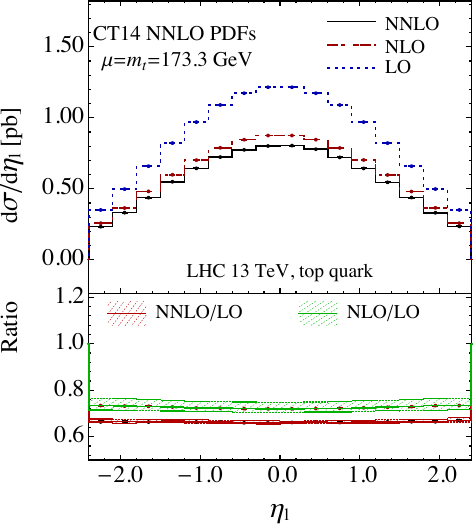}
  \hspace{0.4in}
  \includegraphics[width=0.4\textwidth]{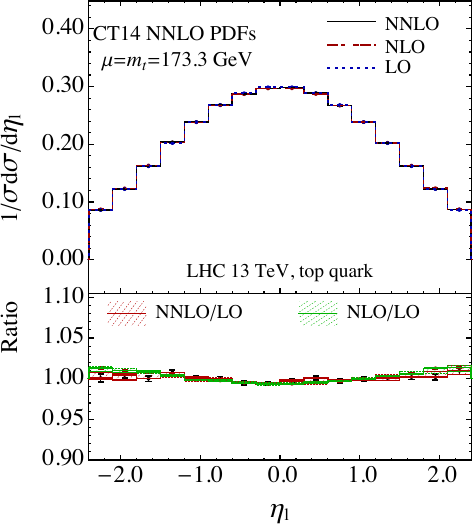}
  \end{center}
  \vspace{2ex}
  \caption{\label{fig:fid1}
   Predicted pseudo-rapidity distribution of the charged lepton 
   from $t$-channel single top-quark production and decay at 13 TeV
   after fiducial cuts,
   including full corrections, with and without normalization respectively.}
\end{figure}

\begin{figure}[h!]
  \begin{center}
  \includegraphics[width=0.4\textwidth]{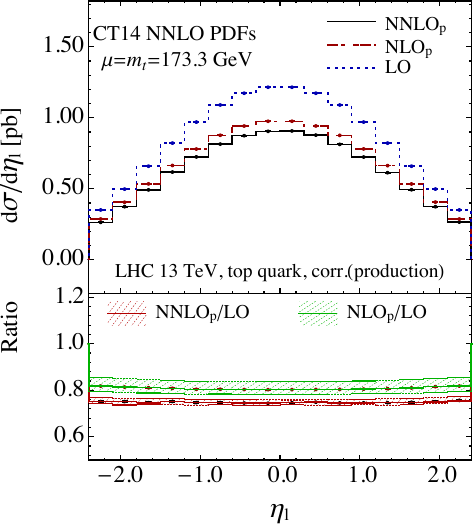}
  \hspace{0.4in}
  \includegraphics[width=0.4\textwidth]{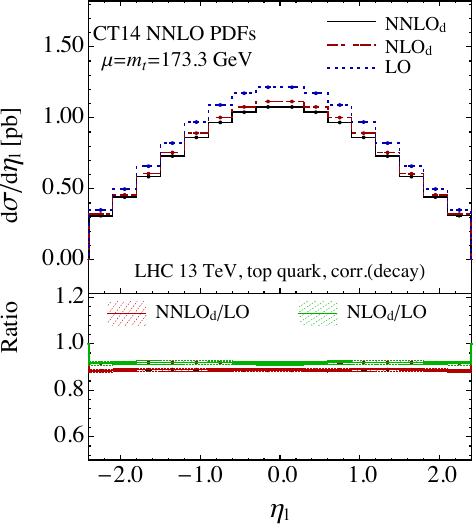}
  \end{center}
  \vspace{2ex}
  \caption{\label{fig:fid2}
   Predicted pseudo-rapidity distribution of the charged lepton
   from $t$-channel single top-quark production and decay at 13 TeV
   after fiducial cuts,
   including corrections from production and decay, respectively.}
\end{figure}

Predicted kinematic distributions within the fiducial volume can be used to compare directly with 
measurements without unfolding procedures.  In Fig.~\ref{fig:fid1} we plot the pseudo-rapidity distribution
of the charged lepton, without and with normalization to the total rate for
top quark production.  The QCD corrections are almost constant over 
the full range for the unnormalized distribution. The NNLO corrections
are about -6\% and reduce the scale variations significantly. We 
observe the large gaps between the NLO and NNLO error bands.  For the normalized
distributions the QCD corrections are small and within 1\% in general.
In the lower panel of the plot on the normalized distribution, the NNLO
results show MC integration fluctuations at the level of a few per mil, also 
shown by the error bars. In Fig.~\ref{fig:fid2}
we show results for the same unnormalized distribution but with QCD
corrections only from the production or from the decay.  Both corrections show
little dependence on the pseudo-rapidity, just as for the full corrections.
The size of the corrections from decay are about one-half those from production.

%%%%%%%%%%%%% begin figure 14 %%%%%%%%%%%%%%%%
%%%%%%%%%%%% end   figure 14 %%%%%%%%%%%%%%%%

%%%%%%%%%%%%% begin figure 15 %%%%%%%%%%%%%%%%
%%%%%%%%%%%% end   figure 15 %%%%%%%%%%%%%%%%

\begin{figure}[h!]
  \begin{center}
  \includegraphics[width=0.4\textwidth]{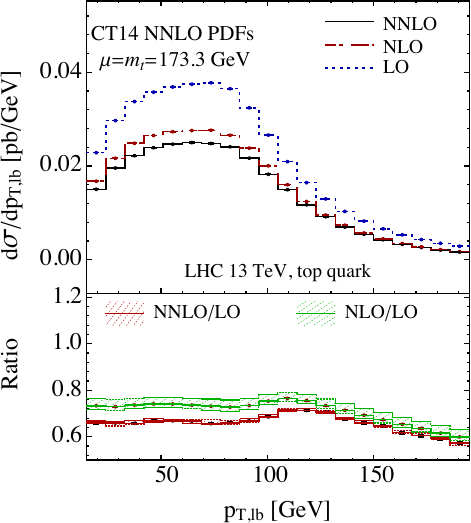}
  \hspace{0.4in}
  \includegraphics[width=0.413\textwidth]{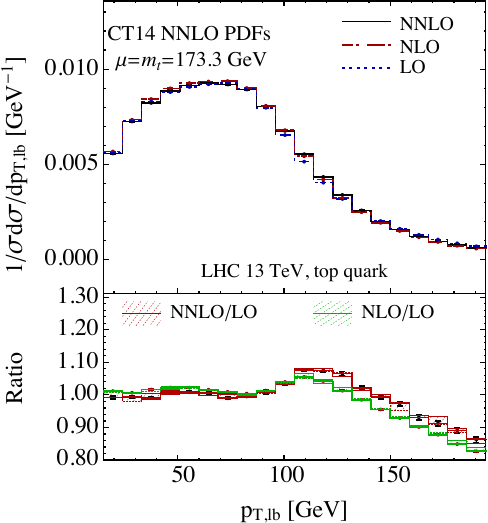}
  \end{center}
  \vspace{2ex}
  \caption{\label{fig:fid3}
   Predicted transverse momentum distribution of the charged lepton and
   $b$-jet system from $t$-channel single top-quark production at 13 TeV
   after fiducial cuts,
   with full corrections included, with and without normalization respectively.}
\end{figure}

\begin{figure}[h!]
  \begin{center}
  \includegraphics[width=0.4\textwidth]{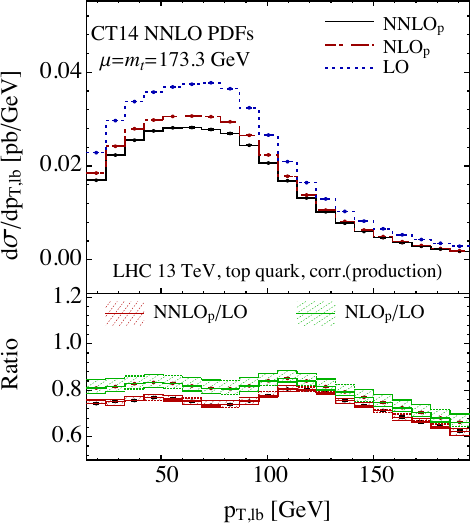}
  \hspace{0.4in}
  \includegraphics[width=0.4\textwidth]{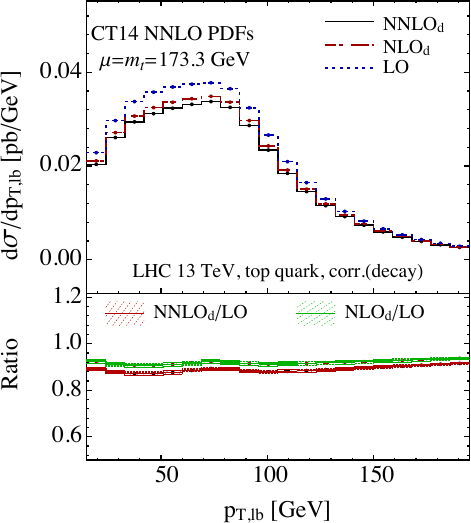}
  \end{center}
  \vspace{2ex}
  \caption{\label{fig:fid4}
   Predicted transverse momentum distribution of the charged lepton and
   $b$-jet system from $t$-channel single top-quark production at 13 TeV
   after fiducial cuts,
   including corrections from production and decay respectively.}
\end{figure}

In Figs.~\ref{fig:fid3} and~\ref{fig:fid4} we show similar results 
for the transverse momentum distribution of the system composed of the 
charged lepton and $b$-jet.  The QCD corrections from decay do not change the
shape of the distribution. The size of the corrections from production
tend to be enhanced in the tail region. The QCD corrections
induce a nontrivial change in the shape of the normalized
distribution, as can be seen from the plot of the right side of Fig.~\ref{fig:fid3}.
The NNLO corrections can reach 5\% in the tail region.   

\begin{figure}[h!]
  \begin{center}
  \includegraphics[width=0.4\textwidth]{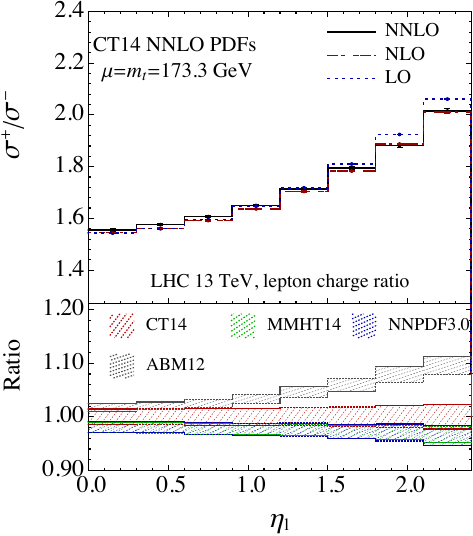}
  \hspace{0.4in}
  \includegraphics[width=0.41\textwidth]{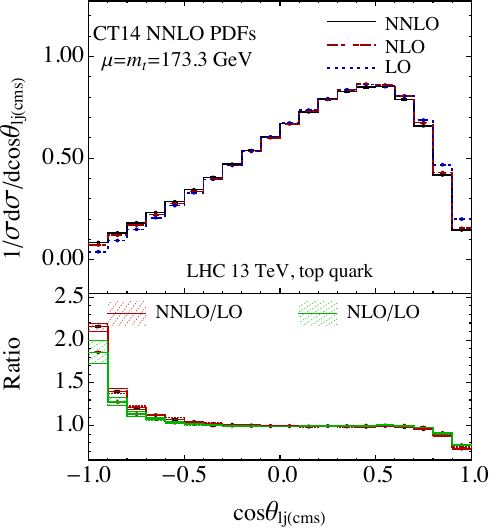}
  \end{center}
  \vspace{2ex}
  \caption{\label{fig:fid5}
   Predicted lepton charge ratio as a function of the pseudo-rapidity (left plot),
   and normalized angular distribution between the charged lepton and
   the non-$b$ jet in the rest frame of the top quark (right plot), 
   from $t$-channel single top-quark production at 13 TeV
   after fiducial cuts, including full corrections.}
\end{figure}

In the plot on the left of Fig.~\ref{fig:fid5} we show the predicted lepton charge
ratio as a function of the pseudo-rapidity. It is sensitive to the ratio 
$u/d$ of PDFs at different momentum fractions. Predictions at different orders 
in the upper panel are all based on the CT14
NNLO PDFs.   The ratio increases with the pseudo-rapidity
because the $u$-valence PDF is harder, extending into the region of 
higher $x$ than the $d$ valence PDF, where the sea-quark contributions
are also smaller. 
%The ratio of the NLO and NNLO predictions to the
%LO ones are shown in the lower panel by lines with error bars
%from MC statistics.  
The QCD corrections are small in general. The NNLO
corrections are within 1\% over the entire kinematic range.
There are four hatched bands in the lower panel representing the spread
of the LO predictions from CT14~\cite{Dulat:2015mca},
MMHT2014~\cite{Harland-Lang:2014zoa}, NNPDF3.0~\cite{Ball:2014uwa},
and ABM12~\cite{Alekhin:2013nda} PDFs with
individual $1\,\sigma$ PDF uncertainties.
Similar to the inclusive charge ratio shown previously, the dependence
of the ratio on PDFs is much larger than the size of QCD corrections.   
These results indicate that sufficiently precise experimental measurements of 
the lepton charge ratio will further constrain the PDFs without
much effect from perturbative uncertainties. 

In the plot on the right side of Fig.~\ref{fig:fid5} we show a normalized angular
distribution in top quark production. The angle $\theta$ is defined in the reconstructed top 
quark rest frame between the charged lepton and the non-$b$ jet.   This type of 
distribution is used typically for measurements of the top-quark polarization.  At LO the 
top quark is produced highly
polarized along direction of the {\it spectator quark}. 
\footnote{We might remark that a {\it spectator quark} is not well defined
at higher orders in QCD since there is additional radiation which is 
indistinguishable from the light quark initiated from the EW vertex.}
Ideally one should see almost a straight line from 0 to 1 as $\cos \theta$ is varied from $-1$ to
$+1$.  Acceptance affects the distribution in the forward region since the charged
lepton tends to be soft there.  The QCD corrections
can be large at both forward and backward angles, as can be seen in the lower panel.  The 
conventional forward-backward asymmetry of the angular
distribution is proportional to the top-quark polarization.  The
predictions are 0.383, 0.362, and 0.346 at LO, NLO, and NNLO, respectively.
Thus the NNLO correction is about -4\% on the forward-backward asymmetry.        

%%%%%%%%%%%%% begin figure 16 %%%%%%%%%%%%%%%%
%%%%%%%%%%%% end   figure 16 %%%%%%%%%%%%%%%%

%%%%%%%%%%%%% begin figure 17 %%%%%%%%%%%%%%%%
%%%%%%%%%%%% end   figure 17 %%%%%%%%%%%%%%%%

%%%%%%%%%%%%% begin figure 18 %%%%%%%%%%%%%%%%
%%%%%%%%%%%% end   figure 18 %%%%%%%%%%%%%%%%

\section{Summary}
\label{sec:conclusions}
We presented a detailed phenomenological study of the
next-to-next-to-leading order QCD corrections for $t$-channel single top
(anti-)quark production including its semi-leptonic
decay at the LHC. The calculations are carried out under the on-shell
top-quark and the structure-function approximations, allowing the QCD corrections 
to be factored into three simpler pieces. 
The neglected corrections are suppressed either by the width
of the top quark or by a color factor of $1/N_c^2$.

The NNLO corrections are generally about $-3\%$ for the
total inclusive rates at LHC with different center of mass energies. The
NNLO corrections can be much larger for differential distributions.  
They can reach a level of $10\%$ or more in certain regions of the transverse
momentum distributions of the top (anti-)quark and the pseudo-rapidity 
distributions of the leading jet. In all cases the scale variations are
greatly reduced by the NNLO corrections. We also show a comparison of
the normalized parton-level distributions to the recent data from the ATLAS
8 TeV measurement. The NNLO corrections tend to move theoretical predictions
closer to the measured transverse momentum distribution of the top
(anti)-quark, though the reported 8 TeV data still have large uncertainties.

After top quark decay is included, we calculate and present cross sections in a restricted 
fiducial volume at 13 TeV, approximating experimental selections.  The QCD corrections 
are more pronounced in this case.
The NNLO corrections are about $-6\%$ 
for the total rate and similarly for the
kinematic distributions in the fiducial volume. The corrections from pure decay are
generally half the size of corrections from pure production. For normalized distributions
the QCD corrections are small in general.  Our predictions of the lepton
charge ratio are stable against QCD corrections.  Experimental measurements of 
this ratio can potentially provide further constraints on the ratio of the $u/d$ 
parton distributions in the proton.  Lastly we point out
that the NNLO QCD corrections can induce about $6\%$ shift on the acceptance defined 
as ratio of fiducial to inclusive cross sections.  They may have effects
of a similar level for the unfolded inclusive cross sections in experimental
measurements, which are used for extraction of the electroweak coupling strength.
Further studies are required to refine the exact effects of the NNLO QCD
corrections on the acceptance used in experimental analyses. These can include
a detailed comparison of the NNLO QCD predictions with the NLO predictions
matched with parton showering, or even a possible match of the NNLO predictions
with parton showering.  Phenomenologcal studies such as these and detailed 
comparisons with data are left for future work.

\begin{acknowledgments}
ELB's work at Argonne is supported in part by the U.S. Department of Energy under 
Contract No. DE-AC02-06CH11357.    
The work of JG is sponsored by the Shanghai Pujiang Program.
H.X.Z was supported in part by the Office of Nuclear Physics of the U.S. Department of Energy under Contract No. DE-SC0011090.
We thank K. Melnikov and F. Caola for private communications on aspects of 
their NNLO calculations.  We acknowledge valuable conversations with T. Gehrmann, 
P. Nadolsky, A. Papanastasiou,  A. Signer, Z. Sullivan, and C.-P. Yuan.
We thank Southern Methodist University for
the use of the High Performance Computing facility ManeFrame. \\
\end{acknowledgments}

\bibliography{tchannellv}
\bibliographystyle{jhep}

\end{document}